\documentclass[journal]{IEEEtran}

\usepackage{amsmath,graphicx}
\usepackage{cite}
\usepackage{amssymb,amsfonts}
\usepackage{algorithmic}
\usepackage{textcomp}
\usepackage{xcolor}
\usepackage{stfloats}
\usepackage{booktabs}  
\usepackage{threeparttable}
\usepackage{multirow} 
\usepackage{epsfig}
\usepackage{threeparttable}
\usepackage{amssymb}
\usepackage{verbatim}
\usepackage{subfigure}
\usepackage{graphicx}
\usepackage{float}
\usepackage{color}
\usepackage{stfloats}
\usepackage{flushend}
\usepackage{url}
\usepackage[linesnumbered,boxed,ruled,commentsnumbered]{algorithm2e}

\ifCLASSINFOpdf
\else
\fi
\begin{document}
%
\title{CNS-Net: Conservative Novelty Synthesizing Network for Malware Recognition in an Open-set Scenario}
%
%
%

\author{Jingcai~Guo,
        Song~Guo,
        Shiheng~Ma, 
        Yuxia~Sun, 
        and~Yuanyuan~Xu

\thanks{J. Guo and S. Guo are with the Department of Computing, The Hong Kong Polytechnic University, Hong Kong SAR, China}
\thanks{S. Ma is with the Department of Computer Science and Engineering, Shanghai Jiao Tong University, Shanghai 200030, China.}
\thanks{Y. Sun is with the Department of Computer Science, Jinan University, Guangzhou 510632, China.}
\thanks{Y. Xu is with the School of Information and Software Engineering, University of Electronic Science and Technology of China, Chengdu 610054, China.}
} 

\maketitle

\begin{abstract}
We study the challenging task of malware recognition on both known and novel unknown malware families, called malware open-set recognition (MOSR). 
Previous works usually assume the malware families are known to the classifier in a close-set scenario, i.e., testing families are the subset or at most identical to training families. 
However, novel unknown malware families frequently emerge in real-world applications, and as such, require to recognize malware instances in an open-set scenario, i.e., some unknown families are also included in the test-set, which has been rarely and non-thoroughly investigated in the cyber-security domain. 
One practical solution for MOSR may consider jointly classifying known and detecting unknown malware families by a single classifier (e.g., neural network) from the variance of the predicted probability distribution on known families. 
However, conventional well-trained classifiers usually tend to obtain overly high recognition probabilities in the outputs, especially when the instance feature distributions are similar to each other, e.g., unknown v.s. known malware families, and thus dramatically degrades the recognition on novel unknown malware families. 
To address the problem and construct an applicable MOSR system, we propose a novel model that can conservatively synthesize malware instances to mimic unknown malware families and support a more robust training of the classifier. 
More specifically, we build upon the generative adversarial networks (GANs) to explore and obtain marginal malware instances that are close to known families while falling into mimical unknown ones to guide the classifier to lower and flatten the recognition probabilities of unknown families and relatively raise that of known ones to rectify the performance of classification and detection. A cooperative training scheme involving the classification, synthesizing and rectification are further constructed to facilitate the training and jointly improve the model performance. 
Moreover, we also build a new large-scale malware dataset, named MAL-100, to fill the gap of lacking large open-set malware benchmark dataset. Experimental results on two widely used malware datasets and our MAL-100 demonstrate the effectiveness of our model compared with other representative methods.
\end{abstract}


%
\IEEEpeerreviewmaketitle

\section{Introduction}

\begin{figure}[t]
\centering
\includegraphics[width=0.40\textwidth]{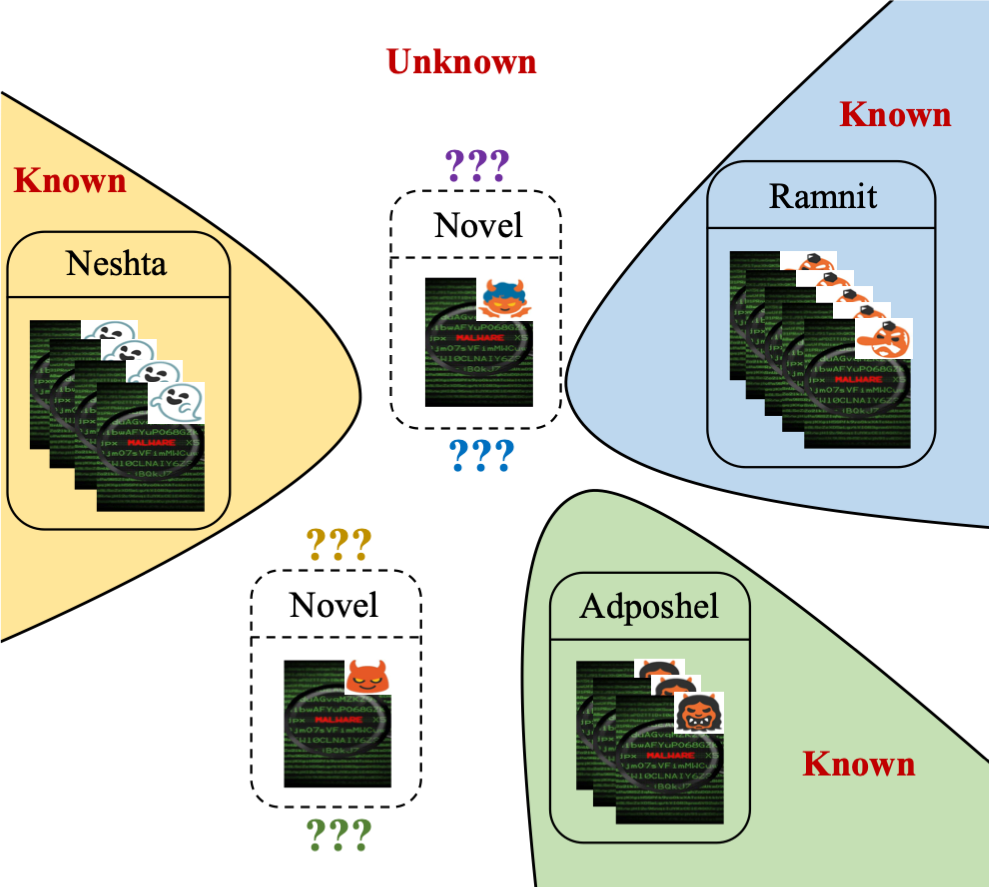}
\caption{Malware recognition in the open-set scenario. The instances of only known families, i.e., denoted as solid boxes, are presented during the training phase, while the model is expected to recognize instances from both known and unknown (dashed boxes) families, respectively (better viewed in color).}
\label{openset}
\end{figure}

\IEEEPARstart{M}{alicious} software, a.k.a. malware, includes computer viruses, spyware, trojan horses, worms, etc., can cause severe damages to various devices and public networks and result in many issues in cyber-security. In the past few years, the machine deep learning has made tremendous success across multiple real-world applications~\cite{deng2009imagenet,batista2004study,lin2011machine,guo2020novel,lv2014traffic,nguyen2015topic,guo2020dual,lian2018xdeepfm,he2016deep,simonyan2014very,guo2019adaptive,yang2015facial,shone2018deep,mahapatra2021medical,guo2023graph,liu2022towards,guo2021conservative,lu2023decomposed,wang2023data,liu2023zsl,guo2023application,huo2023offline,zhou2021device,guo2021learning,wang2022exploring,ma2019position,guo2016improved,huo2022procc,zhou2022cadm,guo2019ams,wang2022efficient,guo2019ee,zhou2021octo,guo2022fed}. Recently, the malware instances have been increasing over the years and brought many challenges \cite{rieck2008learning,zhou2012dissecting}. The malware recognition aims at classifying numerous malware instances into different families, i.e., a group of malware instances with similar attack techniques, and then further investigations and precautionary measures can be made. Previous works of malware recognition usually hold a relatively strong assumption that all malware families are known to the recognition system, which means that the testing instances belong to the same families with training instances in a close-set scenario. 
This setting is partially acceptable because on the one hand, the malware families are relatively stable for a certain period of time, and on the other hand, to fully collect all malware families from the whole cyber networks is impossible. Therefore, the close-set malware recognition has been extensively investigated in the past few years  \cite{dahl2013large,kong2013discriminant,pascanu2015malware,huang2016mtnet}. 

However, with the recent popularity of network applications, more and more malware attackers are constantly releasing malware instances belonging to various known families and many more novel unknown families. As pointed by NortonLifeLock (formerly known as Symantec) \cite{NortonLifeLock}, over 317 million new malware instances are uncovered annually, and many of them are not belong to any malware family we have already known before. These instances of novel unknown malware families differ in several features such as statistical characteristics, attack techniques, and so on. Under such a situation, conventional malware recognition systems fail to handle the recognition task requiring not only classifying known families but also detecting novel unknown families at the same time. As demonstrated in Fig. \ref{openset}, instances from known malware families, e.g., ``Neshta'', ``Ramnit'' and ``Adposhel'', are used in the training phase of the classifier. During testing or inference, the classifier is expected to first correctly distinguish whether an instance is from these known families, and then to classify it into a specific known family as accurately as possible. This task can be considered as the malware open-set recognition (MOSR), where the concept of ``open-set'' is noticed in several recent works of computer vision domain \cite{scheirer2012toward}. 

Being an important and practical real-world application, the malware recognition in the open-set scenario has been rarely investigated in the cyber-security domain and hinders the further development of malware recognition systems. 
Inspired by several works from computer vision domain, the open-set image recognition can be achieved by a single classifier (e.g., neural network) and determined by the variance of the network outputs. For example, the output softmax \cite{liu2016large} of a multi-class classification network can represent the predicted recognition probability distribution on known families. Thus, the classification of known classes can be determined by the maximum dimensional value of the predicted recognition probability, and the detection of novel unknown classes can be determined by a threshold probability compared to all dimensional values, i.e., if all the dimensional values of a testing instance are smaller than a threshold probability, the instance is considered to be from a novel unknown class. 
Such a recognition framework works well in the computer vision domain \cite{bendale2016towards,ge2017generative,neal2018open,liu2019large}. However, the difference in the characteristic features of malware instances is far less than that of images and thus can result in many overlaps among different malware families. 
This difference may make the framework not applicable to the MOSR task because the predicted recognition probability may tend to be overly high on all malware instances from both known and unknown malware families.

To deal with these problems and construct an applicable MOSR system, we propose the conservative novelty synthesizing network (CNSNet) to coordinate and support the malware recognition system to fit the open-set scenario. Specifically, we make use of the generative adversarial networks (GANs) to synthesize several marginal malware instances that are close to known families while not belong to any of them. Such synthesized instances are then be assigned as the mimical novel unknown malware families and implicitly rectify the classifier to be relatively more sensitive to known families while significantly suppress the sensitivity to unknown ones at the same time. This rectification can be achieved by constraining two regularizers on the synthesized instances that consider lowering and flattening the recognition probabilities in a global view (overall unknown probabilities flattening), and minimizing the batch-level recognition probabilities in a local view (specific known families exclusion), respectively. As such, our model is able to better distinguish between known and unknown malware families and improve the classification and detection performance. To jointly optimize the classification, synthesizing, and rectification in a unified framework, we further construct a cooperative training scheme that allows each component to complement each other and improve alternately. 
Moreover, to verify our model works well in general, we also paid significant effort to build and propose a new large-scale malware dataset containing more than 50 thousand malware instances from 100 malware families, termed as MAL-100, to fill the gap of lacking large malware open-set benchmark dataset in malware recognition domain. Experimental results verified the effectiveness of our proposed method and demonstrated the flexibility of our proposed large-scale malware dataset.

 In summary, our contributions are four-fold.
 \begin{itemize}
 	\item We present the first formal investigation for malware recognition in the open-set scenario.
 	\item We propose a novel malware open-set recognition framework that can conservatively synthesize marginal malware instances to mimic novel unknown families and jointly improve the performance of classification and detection.
 	\item We propose a cooperative training scheme to unify the system objective and facilitate the training process.
 	\item We propose a large-scale malware benchmark dataset, namely MAL-100, to complement the malware recognition in the open-set scenario, which can continuously contribute to future research.
 \end{itemize}
 
 The rest of this paper is organized as follows. Section \ref{related_work} introduces the related work. Then, in Section \ref{method}, we present our proposed method. Section \ref{experiment} discusses the experiment, and the conclusion and future work are detailed in Section \ref{conclusion}.

\section{Related Work}
\label{related_work}

\subsection{Malware Family Recognition}
Conventional malware recognition systems aim at classifying malware instances into several known families in a close-set scenario, mainly based on malware characteristic features collected by analyzing instances in a static or dynamic fashion \cite{kalash2018malware,sun2018deep}. Since the dynamic analysis of malware is usually time-consuming and requires numerous extra sand-box experiments on instances, which makes real-time recognition impossible in real-world applications. Thus, the static analysis is more appropriate for large-scale malware recognition systems and be adopted by most existing works. The static analysis can involve several static characteristic features of instances such as activation mechanisms, installation methods, natures of carried malicious payloads and so on \cite{zhou2012dissecting, yerima2018droidfusion}. 

Recent advanced close-set malware recognition works usually employ classic machine leaning (CML) and deep leaning (DL) techniques to implement their models. 
For the CML-based methods, Wu \textit{et al.} \cite{wu2012droidmat} proposed to use k-means and k-nearest neighbor (KNN) algorithms to classify malware apps based on static features such as intents, permissions, and API calls. 
Dahl \textit{et al.} \cite{dahl2013large} applied random projections to do the dimension reduction on the original input space and train several large-scale neural network systems to recognize malware instances. 
Kong \textit{et al.} \cite{kong2013discriminant} proposed to discriminant malware distance metrics that evaluate the similarity between the attributed function call graphs of two malware programs. 
Arp \textit{et al.} \cite{arp2014drebin} proposed to implement an on-device recognition model based on support vector machine (SVM) considering network access, API calls, permissions, etc. 
Wang \textit{et al.} \cite{wang2014exploring} evaluated the usefulness of risky permissions for malware recognition using SVM, random forest, and decision trees. 
Wang \textit{et al.} \cite{wang2017characterizing} focused on detecting malicious apps by using linear SVM, logistic regression, random forest, and decision tree on features of static analysis. 
Fan \textit{et al.} \cite{fan2017dapasa} proposed to make use of sensitive subgraphs to construct five features then fed into several machine learning algorithms such as decision tree, random forest, KNN, and PART for detecting Android malware piggybacked apps in a binary fashion. Yerima \textit{et al.} \cite{yerima2019droidfusion} proposed to train and fuse multi-level classifiers to form a final strong classifier. 

Most recently, several methods take a further step towards accurately recognizing malware instances by deep leaning techniques. 
Pascanu \textit{et al.} \cite{pascanu2015malware} proposed to use echo state networks and recurrent neural networks (RNNs) to better extract features from executed instructions, robust and time domains to better recognize malware attacks. 
Huang \textit{et al.} \cite{huang2016mtnet} proposed a novel multi-task deep learning architecture for malware classification on the binary malware classification task. 
Ni \textit{et al.} \cite{ni2018malware} converted the disassembled malware features into gray images and used the convolutional neural network (CNN) to recognize malware families. 
Vasan \textit{et al.} \cite{vasan2020imcfn} proposed to apply the assemble and fine-tuned CNN architectures to capture more semantic and rich features based on image representation of malware instances. 

Despite the progress made, however, conventional malware recognition systems are usually conducted in a close-set scenario that only known malware families can be handled, and lack the ability to deal with novel unknown malware families at the same time. Recently, as increasing novel malware families have been constantly released, an open-set malware recognition system is urgently needed in real-world applications. In this paper, we focus on formally investigating the malware recognition problem in the open-set scenario.

\subsection{Image Open-set Recognition}
The open-set recognition was first investigated in the area of computer vision on images \cite{scheirer2012toward} where the methods can be roughly grouped into three main-streams including classic machine leaning (CML)-based methods, deep leaning (DL)-based methods, and extreme value theory (EVT)-based methods.

Some early works are mainly implemented with several classic machine learning algorithms such as support vector machine (SVM), nearest neighbors, sparse representation (SR), etc. The 1-vs-Set Machine \cite{scheirer2012toward} and the W-SVM \cite{scheirer2014probability} applied the SVM to achieve the open-set detection. The former proposed to sculpt a decision space from the marginal distances of a one-class or binary SVM with a linear kernel to detect, and the latter combined the SVM with the useful properties of statistical extreme value theory to calibrate the classifiers. SSVM \cite{junior2016specialized} proposed to balance the empirical risk and the risk of the unknowns, which ensuring a finite risk of the unknown classes. OSNN \cite{junior2017nearest} extended the nearest neighbors classifier to open-set scenario by fully incorporating the ability of recognizing samples belonging to classes that are unknown during training. A generalized SR-based classification algorithm \cite{wright2014robust} was then proposed to make use of the generalized pareto extreme value distribution for open-set image recognition. 

Recently, with the rapid development of deep learning techniques, several works adopt the deep neural network to implement the recognition models. Among them, the OpenMax \cite{bendale2016towards} is one of the pioneers that estimates the probability of an input being from an unknown class by using the deep neural networks (DNNs). Later on, several DL-based methods also combined the open-set recognition with the GANs. For example, G-OpenMax \cite{ge2017generative} extended the OpenMax by generating unknown image samples from the known class data with GANs to augment the training set. Slightly different from the G-OpenMax, Neal \textit{et al.} \cite{neal2018open} proposed to generate image samples that are close to training set yet do not belong to any training class for the augmentation. The OLTR \cite{liu2019large} considered a long-tailed and open-ended distribution and proposed to implement the open-set recognition in such a nature, which makes the model capable of dealing with imbalanced classification, few-shot learning, and open-set recognition in one integrated algorithm.

Additionally, the EVT-based methods are a special category derived from statistics that can deal with modeling of statistical extremes \cite{de2007extreme}. Among them, the EVM \cite{rudd2018extreme} is the most representative one that addressed the gap between other open-set recognition methods where most of them take little to no distribution information of the data into account and lack a strong theoretical foundation.

Being successful in the computer vision domain, existing image open-set recognition models may theoretically transfer to malware recognition in an open-set scenario by adjusting the malware features to mimic image features. However, compared to the computer vision domain, the difference in the characteristic features of various malware instances is far less than images, which may also result in overly high recognition probabilities on novel unknown malware families and cannot distinguish from known families.

\subsection{GANs-based Malware Recognition Models}
In recent year, with the increasing popularity of GANs that involves two neural networks contesting with each other in a zero-sum game framework. Several malware researchers also followed this framework to implement their malware recognition systems. For example, 
tGAN \cite{kim2017malware} proposed to pre-train the GANs with an autoencoder structure to synthesize malware instances based on a small amount of malware data and augment the training volumes. Moreover, the classifier is also transferred from the discriminator of the trained GANs to fully make use of the GANs properties. 
tDCGAN \cite{kim2018zero} followed such an autoencoder pre-train framework and extended tGAN \cite{kim2017malware} to the deep convolutional GANs for better generation and detection abilities on malware instances. 
Differently, Lu \textit{et al.} \cite{lu2019generative} applied the deep convolutional GANs without a pre-train process to synthesize malware instances and augment the training volumes. A ResNet-18 is further trained as the classifier for malware recognition. 
Most recently, Chen \textit{et al.} \cite{chen2021using} proposed to use the GANs to synthesize android malware instances for small malware families and mitigated the problem of instance imbalance malware recognition. 

Our method also implements the recognition system with GANs to synthesize malware instances while differs from existing GANs-based malware recognition models in three aspects including:  
1) Different from most methods that focus on augmenting the training data in known families, we choose to synthesize marginal malware instances that are close to known families while not belong to any of them to mimic the unknown families, and thus we can suppress the classifier sensitivity to unknown families via specific regularizers;  
2) Our synthesizing process is more conservative regarding the synthesizing target where no strong assumption or prior is made on unknown families; 
and 3) Our method focuses on formally investigating the MOSR system that can handle both known and unknown malware families via a single rectified classifier, of which the synthesizing network mainly acts as the supplementary.


\section{Proposed Method}
\label{method}
In this section, we first give the formal problem definition of MOSR. Next, we introduce our proposed method and formulation in detail. More specifically, a synthesizer based on the GANs is trained to synthesize several marginal malware instances to mimic the novel unknown malware families and support the training of the classifier. The classifier is conditioned on lowering and flattening the recognition probabilities of unknown families and relatively raises that of known ones to rectify the performance of classification and detection. The synthesizing and classification networks are jointly optimized to improve each other alternately.

\subsection{Problem Definition}
We start by formalizing the MOSR task and then introduce our proposed method. 
Given a set of labeled known malware family instances $\mathcal{D}=\left \{ x_{i}, y_{i} \right \}_{i=1}^{n}$, where $x_{i}$ is an instance containing several static characteristic features, with family label $y_{i}$ belonging to $k$ known families $\mathcal{F}=\left \{ f_{1}, f_{2}, \cdots , f_{k}\right \}$. The goal is to construct a classifier to recognize malware instances from not only known families $\mathcal{F}$, but also novel unknown families $\mathcal{F}^{u}$ where $\mathcal{F}^{u} \cap \mathcal{F} = \phi$. Specifically, as to the classifier, we can adopt the most widely used cross-entropy loss to supervise the training as:
\begin{equation}
\underset{\omega_{c}}{\min} \ \frac{1}{n} \sum_{i=1}^{n}\sum_{j=1}^{k} -y_{i,j} \log \left ( P_{\omega_{c}}\left (\hat{y_{i,j}} \mid x_{i} \right )\right ),
\end{equation}
where $P_{\omega_{c}}\left ( \cdot \mid \cdot \right )$ is the classifier with trainable parameter $\omega_{c}$, $\hat{y_{i,j}}$ is the probability that the $i$-th instance $x_{i}$ is predicted to be the $j$-th label, and $y_{i,j}$ is the ground-truth, e.g., element within the one-hot form, belonging to known families $\mathcal{F}$. Based on the trained classifier, the malware recognition can be specified as two parallel phases including the classification of known families $\mathcal{F}$, and the detection of novel unknown families $\mathcal{F}^{u}$. Given a testing malware instance $x^{(t)}$. First, the classification can be naturally achieved by the classifier prediction $P_{\omega_{c}}\left (\hat{y}^{(t)} \in \mathcal{F} \mid x^{(t)} \right )$. Next, as to the detection on novel unknown families $\mathcal{F}^{u}$, one possible solution is to consider the predicted probabilities of $x^{(t)}$ on $k$ known families $\mathcal{F}$ as $\left \{ \hat{y}^{(t)}_{i} \right \}_{i=1}^{k} = \left \{ \hat{y}^{(t)}_{1}, \hat{y}^{(t)}_{2}, \cdots , \hat{y}^{(t)}_{k} \right \}$. If all $\hat{y}^{(t)}_{i}$ below a threshold probability, the testing instance $x^{(t)}$ is considered to be from novel unknown families. Such a detection is based on the assumption from image recognition domain that a well-trained classifier $P_{\omega}\left ( \cdot \mid \cdot \right )$ should be familiar with known families $\mathcal{F}$ while unfamiliar with novel unknown families $\mathcal{F}^{u}$. Hence, the predicted probabilities will tend to be high if $x^{(t)}$ comes from known families $\mathcal{F}$, and tend to be low otherwise. 

\begin{figure}[t]
  \centering
  
\subfigure[Class 0]{
   \label{fig:subfig:a} 
   \includegraphics[width=1.4in]{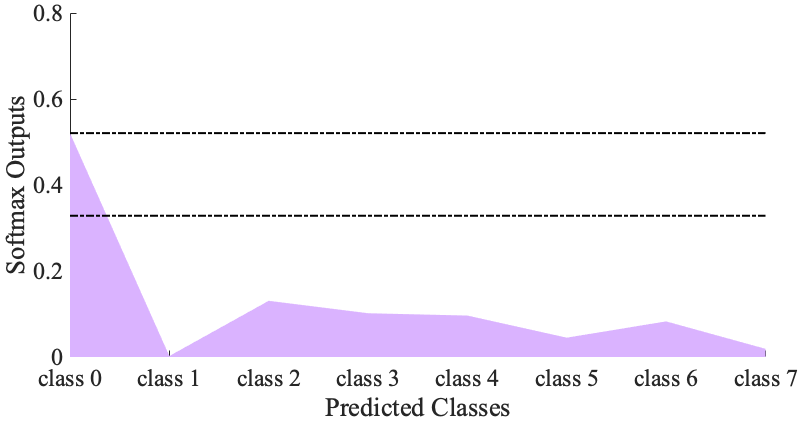}}\hspace{3mm}
\subfigure[Class 1]{
   \label{fig:subfig:b} 
   \includegraphics[width=1.4in]{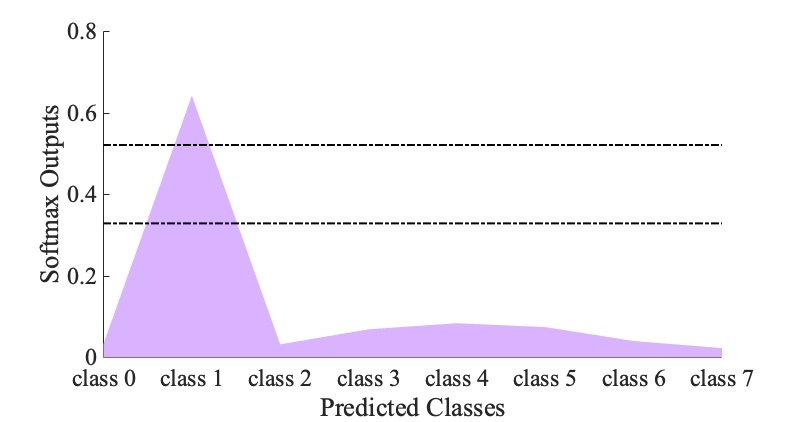}}
   
\subfigure[Class 2]{
   \label{fig:subfig:c} 
   \includegraphics[width=1.4in]{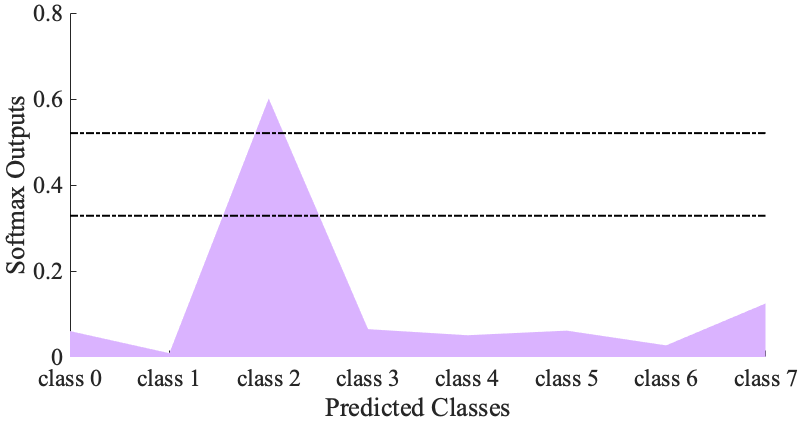}}\hspace{3mm}
\subfigure[Class 3]{
   \label{fig:subfig:d} 
   \includegraphics[width=1.4in]{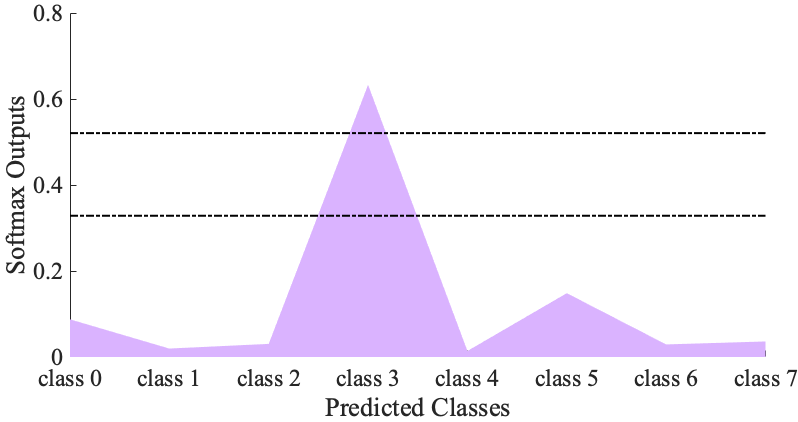}}

\subfigure[Class 4]{
   \label{fig:subfig:e} 
   \includegraphics[width=1.4in]{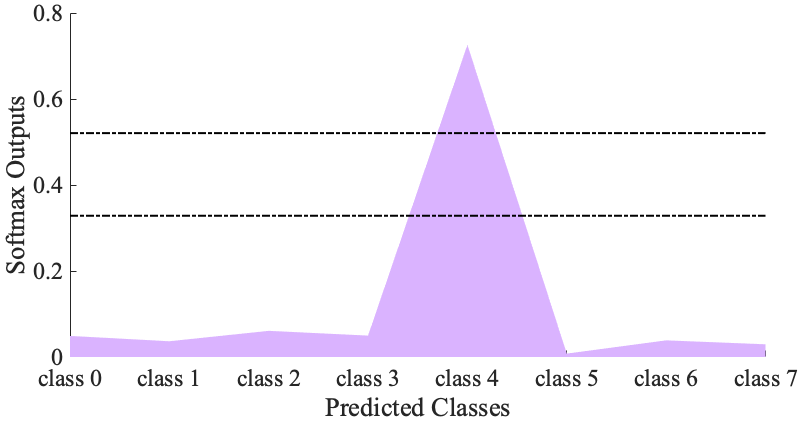}}\hspace{3mm}
\subfigure[Class 5]{
   \label{fig:subfig:f} 
   \includegraphics[width=1.4in]{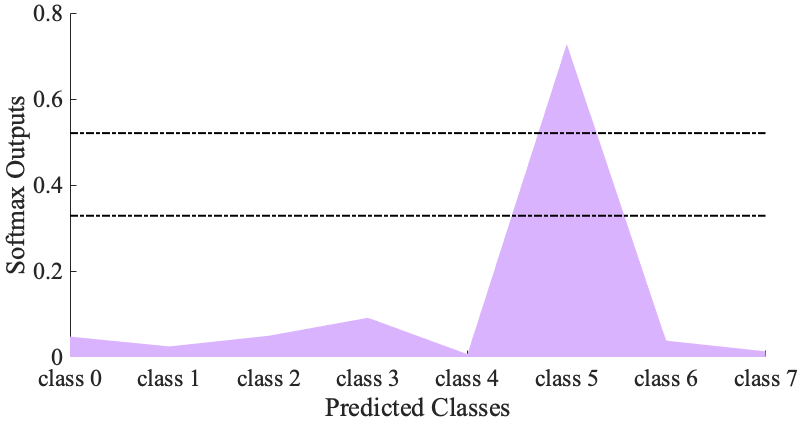}}
   
\subfigure[Class 6]{
   \label{fig:subfig:g} 
   \includegraphics[width=1.4in]{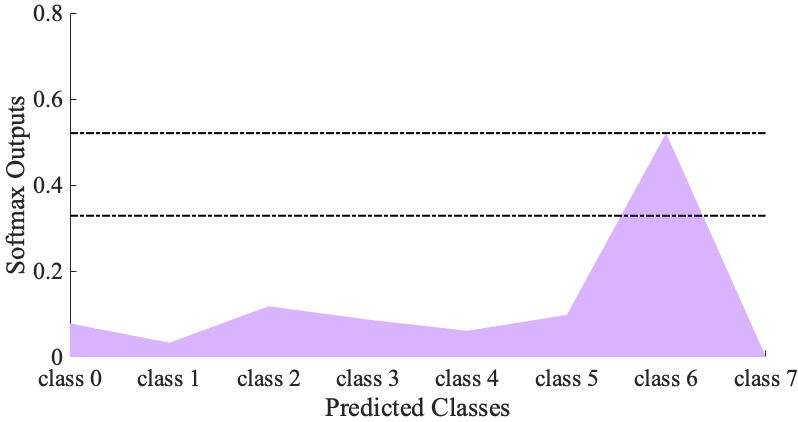}}\hspace{3mm}
\subfigure[Class 7]{
   \label{fig:subfig:h} 
   \includegraphics[width=1.4in]{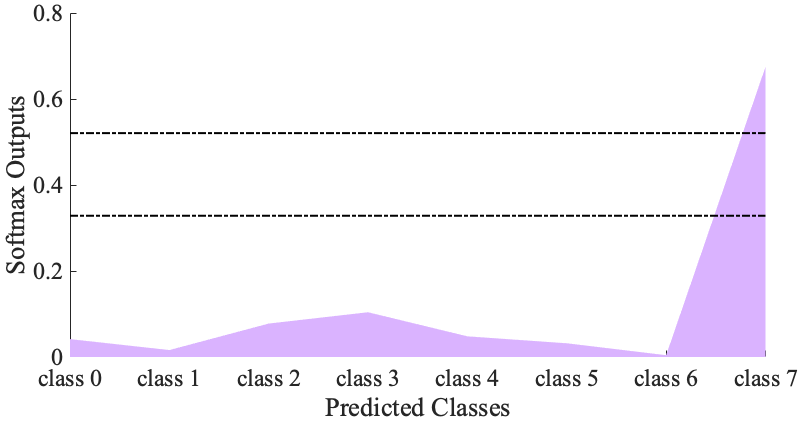}}
   
 \subfigure[Class 8 (unknown)]{
   \label{fig:subfig:i} 
   \includegraphics[width=1.4in]{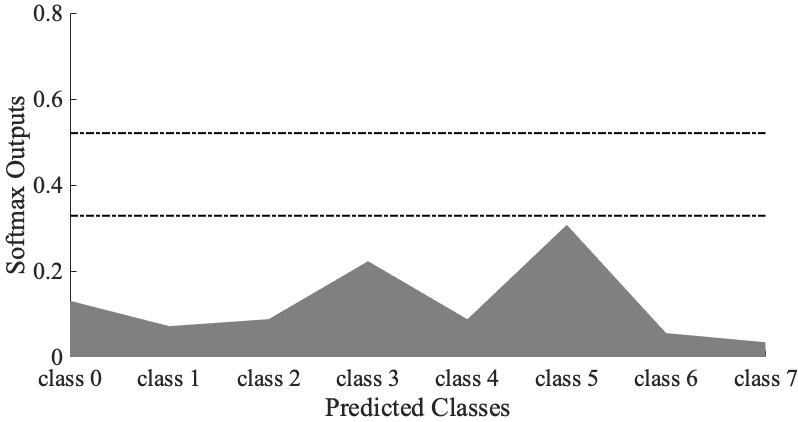}}\hspace{3mm}
\subfigure[Class 9 (unknown)]{
   \label{fig:subfig:j} 
   \includegraphics[width=1.4in]{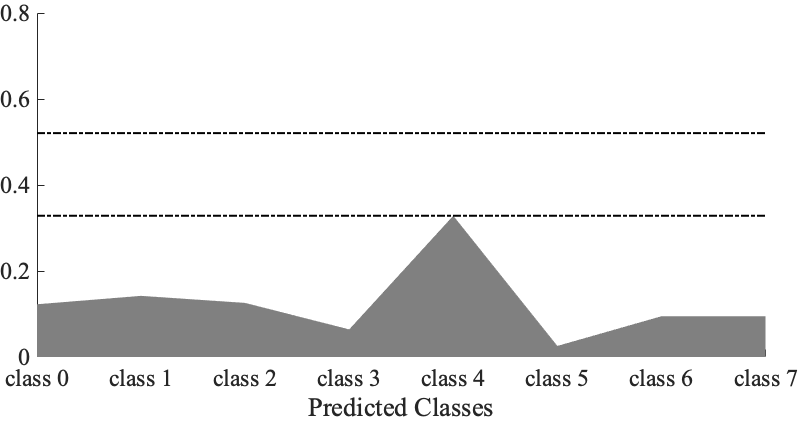}}

\caption{Visualization results of softmax outputs of digits on known classes ``0''$\sim$``7'' and unknown classes ``8'' and ``9''. (a)$\sim$(h) marked as ``violet'' are known classes, and (i), (j) marked as ``gray'' are unknown classes. The two dashed lines denote the upper and lower bounds of the threshold that determine the detection on novel unknown classes (better viewed in color).}
  \label{mnist_prob}
\end{figure}

As demonstrated in Fig. \ref{mnist_prob}, we visualize the softmax outputs (can be approximated as the probability distribution) of digital dataset MNIST \cite{lecun1998mnist} from image domain, where we train a classifier on only eight classes from ``0'' to ``7'' as known classes, and set ``8'' and ``9'' as unknown classes. It can be observed that the testing probability distributions of known classes usually have a relatively high peak, which can be matched to a specific class. While for unknown classes, the testing probability distributions are usually flatter and the peak is obviously low. These natures result in a practical detection by a single classifier, i.e., probability distributions as the recognition confidence with a proper threshold. However, different from the images, the variance of malware families is comparatively small, which means that the testing probability distributions of instances from novel unknown malware families may also have an overly high peak towards known malware families. This difference makes the detection degrades dramatically for malware family recognition in the open-set scenario.

\begin{figure*}[htbp]
\centering
\includegraphics[width=0.77\textwidth]{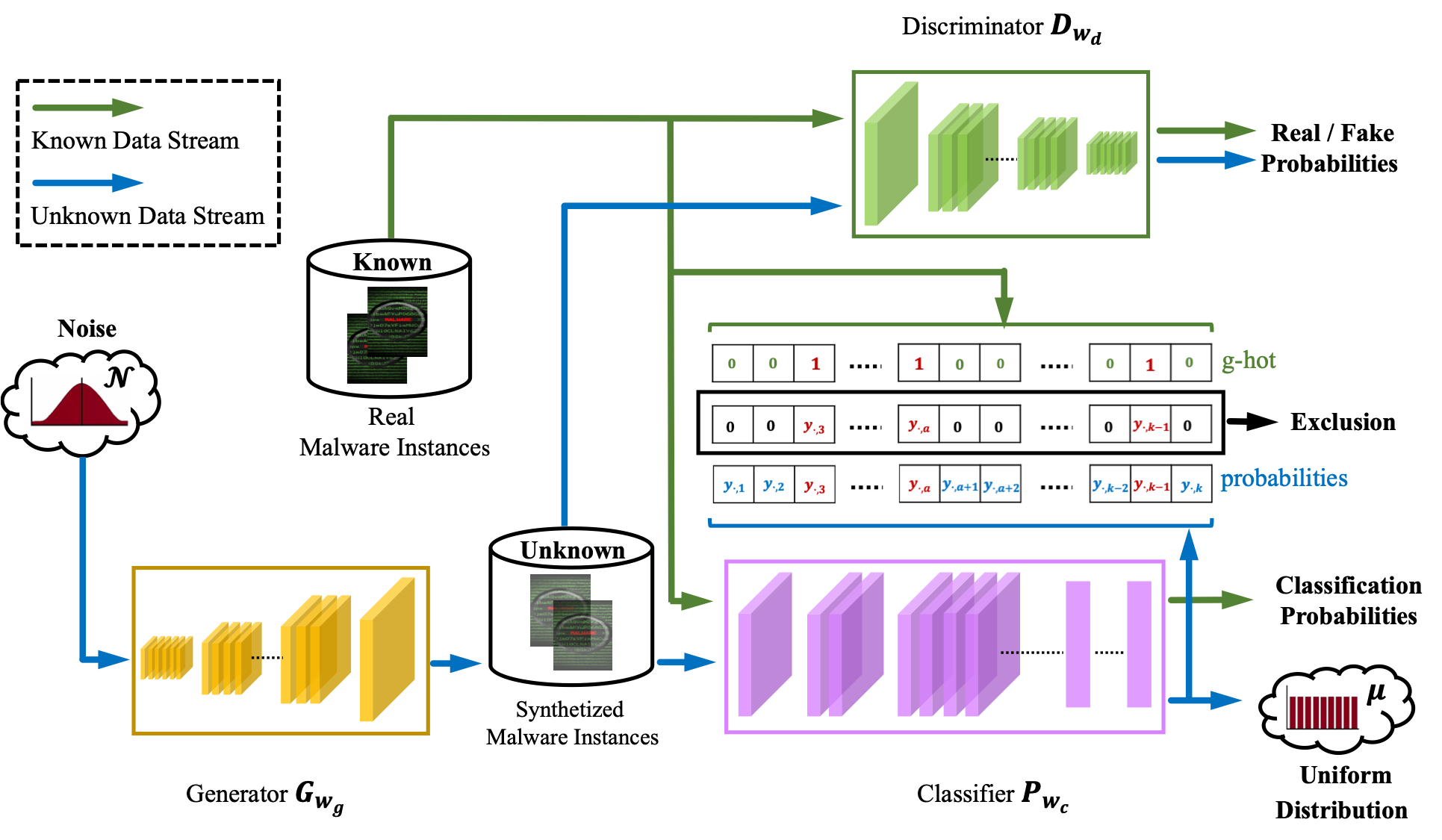}
\caption{The framework of our proposed open-set malware recognition model. It involves a classifier $P_{\omega_{c}}$ and a novelty synthesizing network composed with a generator $G_{\omega_{g}}$ and a discriminator $D_{\omega_{d}}$. The ``green'' data stream denotes known malware families, and the ``blue'' one denotes unknown malware families. The novelty synthesizing network can synthesize marginal malware instances of known families to mimic the novel unknown families, and thus to support the training of the recognition model (better viewed in colour).}
\label{main}
\end{figure*}

\subsection{Conservative Novelty Synthesizing Network}
Conventional methods fail to detect novel unknown malware families because there is no prior information of such families available during the training phase. Worse still, the variance of malware families is usually very small compared to other domains. These deficiencies hinder the feasibility of MOSR system relying on a single classifier. 
To deal with this issue, a practical and straightforward idea is to synthesize several malware instances to mimic novel unknown malware families as the prior, and to support the training of the classifier. 

\subsubsection{Classification Network}
Regardless of the synthesizing process, suppose we have assigned several malware instances of novel unknown families, we can make use of them to support the training phase of the classifier. Intuitively, the classifier should be trained to obtain a probability distribution of an instance from unknown families as flat as possible, which means that the classifier has no judgment regarding the instance belonging to any known families. Thus, we can construct the classifier conditioned on the supervision as:
\begin{equation}
\begin{aligned}
\underset{\omega_{c}}{\min} \ \ &\frac{1}{n} \sum_{1 \leq i \leq n} \sum_{1 \leq j \leq k} -y_{i,j} \log \left ( P_{\omega_{c}}\left (\hat{y_{i,j}} \mid x_{i} \right )\right ) \\
&+ \beta \cdot \frac{1}{m} \sum_{1 \leq l \leq m} \sum_{1 \leq j \leq k} \mathcal{D_{KL}}\left (\mathcal{U}\left ( \tilde{y} \right ) \parallel P_{\omega_{c}}\left (\hat{y_{l,j}} \mid {x_{l}}' \right ) \right )
\end{aligned}
\end{equation}
where $\left \{ {x_{l}}' \right \}_{l=1}^{m}$ are $m$ synthesized malware instances of novel unknown families, classifier $P_{\omega_{c}}\left ( \cdot \mid \cdot \right )$ calculates the probability distributions of ${x_{l}}'$ as $\hat{y_{l,j}}$, and $\mathcal{U}\left ( \tilde{y} \right )$ is the uniform distribution. $\mathcal{D_{KL}}\left ( \cdot \mid \cdot \right )$ calculates the Kullback-Leibler (KL) divergence, a.k.a. relative entropy, and $\beta$ is a hyper-parameter balances the two terms. To minimize this function, we can force the classification results, i.e., probability distributions, of synthesized novel unknown malware families to approximate the uniform distributions. Thus, the obtained classifier will result in the flat probability distributions of them, and can hardly have a clear distinction within known malware families for the instances from novel unknown ones. We call this KL regularizer as the unknown probabilities flattening. 

To flatten the probability distributions of synthesized novel unknown malware families can make the classifier has no judgment regarding the instances belonging to any known families. However, the classifier still lacks the ability to exclude an instance from novel unknown malware families out from from any specific known malware family. For example, suppose a mini-batch $\mathcal{B}$ contains $g$ ($g < k$) known malware families. We could further force the classifier to obtain comparatively low probabilities on these $g$ known families. Thus, the classifier supervision can be further specified as:
\begin{equation}
\begin{aligned}
\underset{\omega_{c}}{\min} \ \ 
&\underset{(i)}{\underbrace{\frac{1}{n} \sum_{1 \leq i \leq n} \sum_{1 \leq j \leq k} -y_{i,j} \log \left ( P_{\omega_{c}}\left (\hat{y_{i,j}} \mid x_{i} \right )\right )}} \\
&+ \underset{(ii)}{\underbrace{\beta \cdot \frac{1}{m} \sum_{1 \leq l \leq m} \sum_{1 \leq j \leq k} \mathcal{D_{KL}}\left (\mathcal{U}\left ( \tilde{y} \right ) \parallel P_{\omega_{c}}\left (\hat{y_{l,j}} \mid {x_{l}}' \right ) \right )}} \\
&+ \underset{(iii)}{\underbrace{\gamma \cdot \sum_{{x}' \mid \forall \mathcal{B} \subseteq \mathcal{D}} \left \| \mathcal{V}_{g-hot} \otimes P_{\omega_{c}}\left (\hat{y} \mid {x}' \right ) \right \|_{2}}}, 
\end{aligned}
\end{equation}
where ${x}' \mid \forall \mathcal{B} \subseteq \mathcal{D}$ in term $(iii)$ of Eq. (3) denotes the synthesized instances during a local mini-batch $\mathcal{B}$, $\mathcal{V}_{g-hot}$ denotes the combination of the one-hot forms of these $g$ families, $\otimes$ is the element-wise multiplication, and $\gamma$ is a balancing hyper-parameter. By minimizing the $L^{2}$-norm, we can make the classifier more sensitive to the judgment of synthesized malware instances do not belong to certain known malware families in a local scenario, e.g., a local mini-batch. 

It should be noted that there seems to have a conflict between terms $(ii)$ and $(iii)$ of Eq. (3), because the former regularizes the flat recognition probabilities to be uniformly distributed, while the latter regularizes a few probabilities to be minimized. However, in our method, term $(ii)$ mainly acts as a global condition to be optimized to make the classifier do not classify synthesized malware instances into any known families. While term $(iii)$ acts as a local condition, e.g., during a local mini-batch, to be optimized to make the classifier can certainly exclude these instances out from several specific known families. These two terms perform together to make the classifier more sensitive to detecting novel unknown malware families.

\subsubsection{Synthesizing Network}
Different from several generative models that synthesize simulated instances for data augmentation in conventional classification problem, or synthesize instances of unseen classes in zero/few-shot learning problem \cite{yan2020semantics,li2019zero}, where the former requires the synthesizing on only known classes and the latter can be provided with several auxiliary information of unknown classes (e.g., semantic descriptions shared by both seen and unseen classes), the conditions of synthesizing novel unknown malware families are relatively limited. 
First, there is no extra auxiliary information of unknown malware families, which makes no supervision for the synthesizing process. Second, the synthesized novel unknown malware families should be different from the known ones, but not too different at the same time. 
Theoretically, any instance that differs from the known family distributions can be regarded as from novel unknown families. Thus, a straightforward rule is to sample instances from different distributions compared to the known ones. However, this rule is not that practical since we cannot sample every different distribution in the instance feature space. 
Further inspired by the support vectors of support vector machines \cite{cauwenberghs2001incremental,guo2016improved}, where the marginal instances usually have better discrimination properties, we can then have a more conservative strategy to synthesize several marginal malware instances that are close to known families while not belong to any of them as the mimical novel unknown malware families. 

Inspired by the GANs \cite{goodfellow2014generative} that involves two neural networks that contest with each other in a zero-sum game framework, to achieve the capability of data generation without explicitly modeling the probability density. We also apply the GANs framework to synthesize the marginal malware instances of novel unknown families. In the GANs networks, a generator $G$ is used to sample a latent variable $z$ from a prior distribution, e.g., a Gaussian $\mathcal{N}$, as the input and to generate an output $G\left ( z \right )$. Meanwhile, a discriminator $D$ is trained to distinguish whether an input $x$ is from a target data distribution by mapping $x$ to a probability ranges in $\left [ 0, 1 \right ]$. The generator aims at synthesizing simulated instances as accurately as possible when freezing $D$:
\begin{equation}
\underset{\omega_{g}}{\min} \ \ \frac{1}{n_{s}}\sum_{1 \leq i \leq n_{s}}\log\left ( 1 - D\left ( G\left ( z_{i} \right ) \right ) \right ),
\end{equation}
where $\omega_{g}$ is the trainable parameter associated with $G$. The discriminator $D$ aims to distinguish the synthesized data out from the real data when freezing $G$:
\begin{equation}
\underset{\omega_{d}}{\min} \ \ \frac{1}{n_{s}}\sum_{1 \leq i \leq n_{s}}\left ( \log\left ( D\left ( x_{i} \right ) \right ) + \log\left ( 1 - D\left ( G\left ( z_{i} \right ) \right ) \right ) \right ), 
\end{equation}
where $\omega_{d}$ is the trainable parameter associated with $D$. These two networks confront each other and the training can be optimized by a min-max objective in a compact form as:
\begin{equation}
\underset{G}{\min} \ \underset{D}{\max} \ \ \mathbb{E}_{x \in \mathit{\mathcal{F}}} \left [ \log D(x) \right ] + \mathbb{E}_{z \in \mathcal{N}} \left [ \log \big(1- D\left (G\left(z\right )\right )\big) \right ].
\end{equation}

Different from synthesizing instances of known families, i.e., to learn the known data distribution, our target is to synthesize novel unknown malware families which are expected to be different from known ones. Thus, we construct the modified objective in a compact form as:
\begin{equation}
\begin{aligned}
\underset{G}{\min} \ \underset{D}{\max} \ \ 
&\underset{(i)}{\underbrace{\mathbb{E}_{x \in \mathcal{F}} \left [ \log D(x) \right ] + \mathbb{E}_{{x}' \in G(z)} \left [ \log \big(1-D({x}')\big) \right ]}} \\
&+ \underset{(ii)}{\underbrace{\beta \cdot \mathbb{E}_{{x}' \in G(z)} \left [\mathcal{D_{KL}} \big ( \mathcal{U}(\tilde{y})\parallel P_{\omega_{c}}(\hat{y} \mid {x}') \big ) \right ]}},
\end{aligned}
\end{equation}
where $P_{\omega_{c}} \left( \cdot \mid \cdot \right)$ is the classifier trained on known malware families. It should be noted that terms $(i)$ and $(ii)$ of Eq. (7) are jointly performed to force the synthesizing network, i.e., the generator $G$, to synthesize marginal malware instances that are close to known families while falling into novel unknown ones. More specifically, term $(i)$ can be considered as the naive GANs supervision that makes the generator $G$ to synthesize similar malware instances with known families. Term $(ii)$ can then force the synthesized malware instances to take a small step from ``being similar'' to ``being different'' with known families. Suppose the classifier $P_{\omega_{c}} \left( \cdot \mid \cdot \right)$ is well-trained with $\omega_{c}$. Under such a condition, the obtained probability distributions $P_{\omega_{c}}(\hat{y} \mid {x}')$ should be similar to the uniform distributions. Thus, the KL divergence of term $(ii)$ should be approximate to 0 for any properly synthesized malware instances. On the other hand, if the synthesized malware instances are too different from the known malware families, i.e., relatively far away from the known margins, the discriminator $D$ will dominate the supervision and makes term $(i)$ overly large. Thus, it can be expected that by jointly minimizing terms $(i)$ and $(ii)$, we can force the generator $G$ to conservatively synthesize marginal malware instances of known families to mimic the novel unknown families.

\subsubsection{Cooperative Training}
There are two main components included in our proposed method: 1) the classification network that aims at producing accurate probability distributions for known malware families and flat and low probability distributions of unknown ones, and 2) the novelty synthesizing network that aims to synthesize marginal malware instances of known families. On the one hand, a well-trained classification network can help to force the novelty synthesizing network to synthesize malware instances from being similar to being different with known families, thus can adjust the naive GANs property. On the other hand, a well-trained novelty synthesizing network can also support the training of the classification network with synthesized malware instances that mimics the novel unknown malware families. Hence, these two components can complement each other and improve the overall performance. 
To take advantage of these properties and facilitate the training, we construct a cooperative training scheme to unify the system objective as:
\begin{equation}
\begin{aligned}
\underset{G}{\min} \ &\underset{D}{\max} \ \underset{\omega_{c}}{\min} \ \
 \underset{(i)}{\underbrace{\mathbb{E}_{(x,y)\in \mathcal{F}} \left [ -y\log P_{\omega_{c}}(\hat{y} \mid x) \right ]}} \\
&+\underset{(ii)}{\underbrace{\gamma \cdot \mathbb{E}_{{x}' \in G\left ( z \right ) \mid \forall \mathcal{B} \subseteq \mathcal{D}} \left \| \mathcal{V}_{g-hot} \otimes P_{\omega_{c}}\left (\hat{y} \mid {x}' \right ) \right \|_{2}}} \\
&+\underset{(iii)}{\underbrace{\beta \cdot \mathbb{E}_{{x}' \in G\left ( z \right )} \mathcal{D_{KL}}\left (\mathcal{U}\left ( \tilde{y} \right ) \parallel P_{\omega_{c}}\left (\hat{y} \mid {x}' \right ) \right )}} \\
&+\underset{(iv)}{\underbrace{\mathbb{E}_{x \in \mathcal{F}} \left [ \log D(x) \right ] + \mathbb{E}_{{x}' \in G(z)} \left [ \log \big(1-D({x}')\big) \right ]}},
\end{aligned}
\end{equation}
where terms $(i)$, $(ii)$, and $(iii)$ form the supervision of the classification network, and terms $(iii)$ and $(iv)$ form the supervision of the novelty synthesizing network. Terms $(ii)$ and $(iii)$ are the rectification regularizers that are shared by both the classification and synthesizing networks.

To cooperatively optimize Eq. (8), we construct an alternate update algorithm. Specifically, we maintain a parameter set $\left \{ \omega_{d}, \omega_{g}, \omega_{c} \right \}$ corresponds to the classifier network and the discriminator $D$ and generator $G$ of the novelty synthesizing network. Similar to the training process of GANs, we update each parameter in turn while freezing the other two parameters. The detailed training process is demonstrated in Algorithm 1.

\subsubsection{Recognition}
Similar to most existing methods, once the model has been trained, we separate the classifier and only apply the classification network to the MOSR system in an open-set scenario. Specifically, given a testing malware instance $x^{(t)}$, the classifier calculates its predicted recognition probability distribution $P_{\omega_{c}}\left (\hat{y}^{(t)} \mid x^{(t)} \right )$ on all known malware families $\mathcal{F}$ as $\left \{ \hat{y}^{(t)}_{i} \right \}_{i=1}^{k}$. Then, the detection is first performed by filtering flat and low probabilities of the testing instance as:
\begin{equation}
DET \left (x^{(t)} \mid \hat{y}^{(t)} \right ) = {\bf 1}\left [ \forall \left ( \hat{y}^{(t)}_{i} \in \hat{y}^{(t)} \right ), \hat{y}^{(t)}_{i} < \theta \right ],  	
\end{equation}
where $\theta$ is a threshold probability and ${\bf 1}\left [ \cdot \right ]$ is an indicator function that takes a value of ``1'' if its argument is true, and ``0'' otherwise. If $x^{(t)}$ is not be detected from novel known malware families, it will then be classified into one specific family of known malware families as:
\begin{equation}
CLS \left (x^{(t)} \mid \hat{y}^{(t)} \right ) = \max\left ( \hat{y}^{(t)}, indexes \right ),	
\end{equation}
where the maximum predicted recognition probability $\hat{y}^{(t)}_{i}$ indicates the corresponding known family (e.g., denoted by the indexes). 

\begin{algorithm}[t]

Initialize discriminator parameter $\omega_{d}$\;
Initialize generator parameter $\omega_{g}$\;
Initialize classifier parameter $\omega_{c}$\;
\Repeat
{\text{convergence}}
{
Update the discriminator $D_{\omega_{d}}$ \\
\For {$\left \{ z:z_{1 \to b} \right \} \in \mathcal{N}\left ( 0,1 \right ), \left \{ x:x_{1 \to b} \right \} \in \mathcal{F}$}
{$L_{D} = \frac{1}{b}\sum_{1 \leq i \leq b}\left [ \log D(x_{i}) + \log (1 - D(G(z_{i}))) \right ]$\; 
$\omega_{d} \leftarrow \omega_{d} - \eta \frac{\partial L_{D}}{\partial \omega_{d}}$.}           

Update the the generator $G_{\omega_{g}}$ \\
\For {$\left \{ z:z_{1 \to b} \right \} \in \mathcal{N}\left ( 0,1 \right )$}
{$L_{G} = \frac{1}{b}\sum_{1 \leq i \leq b} \left [ \log (1 - D(G(z_{i}))) \right ] + \frac{\beta}{b}\sum_{1 \leq i \leq b} \left [\mathcal{D_{KL}} \big ( \mathcal{U}(\tilde{y})\parallel P_{\omega_{c}}(\hat{y_{i}} \mid G(z_{i})) \big ) \right ]$\;
$\omega_{g} \leftarrow \omega_{g} - \eta \frac{\partial L_{D}}{\partial \omega_{g}}$.}  

Update the classifier $P_{\omega_{c}}$\\ 
\For {$\left \{ z:z_{1 \to b} \right \} \in \mathcal{N}\left ( 0,1 \right ), \left \{ x, y:x_{1 \to b}, y_{1 \to b}  \right \} \in \mathcal{F}$}
{$L_{P} = \frac{1}{b} \sum_{1 \leq i \leq b} -y_{i} \log \left ( P_{\omega_{c}}\left (\hat{y_{i}} \mid x_{i} \right )\right ) + \frac{\beta}{b} \sum_{1 \leq i \leq b} \mathcal{D_{KL}}\left (\mathcal{U}\left ( \tilde{y} \right ) \parallel P_{\omega_{c}}\left (\hat{y_{i}} \mid G(z_{i}) \right ) \right ) + \frac{\gamma}{b} \cdot \sum_{1 \leq i \leq b} \left \| \mathcal{V}_{g-hot} \otimes P_{\omega_{c}}\left (\hat{y_{i}} \mid G(z_{i}) \right ) \right \|_{2}$\;
$\omega_{c} \leftarrow \omega_{c} - \eta \frac{\partial L_{P}}{\partial \omega_{c}}$.} 
}
\caption{Cooperative Training}
\end{algorithm}


\section{Experiment}
\label{experiment} 

\subsection{Datasets and Settings}
We conduct the experiments on three malware datasets including Malware Classification Challenge (BIG 2015) \cite{ronen2018microsoft,Kaggle}, Mailing \cite{nataraj2011malware}, and our proposed MAL-100. 

\subsubsection{Malware Classification Challenge (BIG 2015) Dataset}
The BIG 2015 dataset is released by Microsoft. It consists of 10,868 publicly available labeled malware instances from 9 families with binary and disassembly files. 

\subsubsection{Mailing Dataset}
The Mailing dataset contains total 9,339 malware instances covering 25 families. The malware binaries sequences are grouped as 8-bit vectors and then are presented as grayscale images.

\begin{figure}[t]
\label{mal100}
\centering
\includegraphics[width=0.4\textwidth]{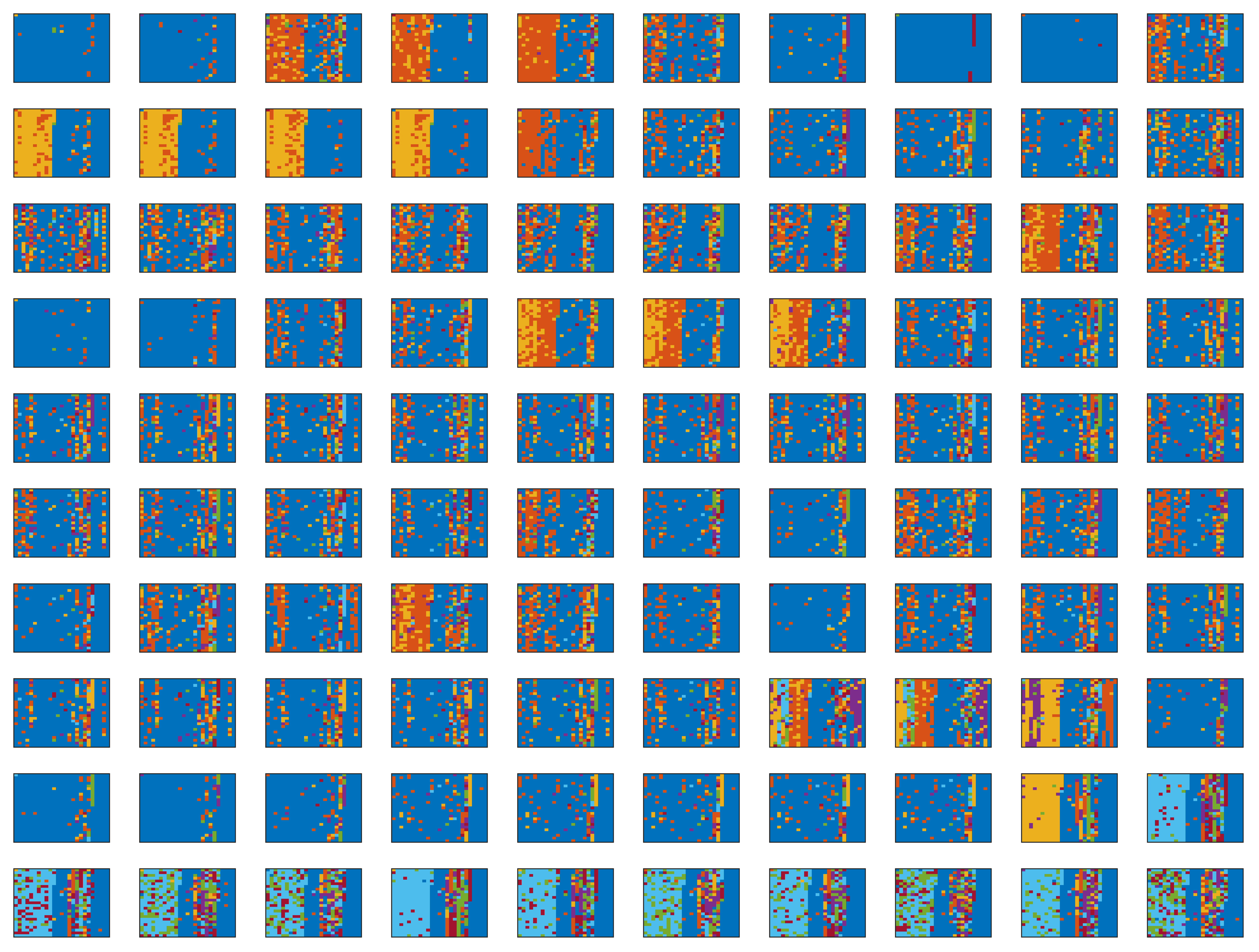}
\caption{Visualization of MAL-100 in 100 mean-instances. Each image in the 10$\times$10 grid denotes a malware family (better viewed in color).}
\end{figure}

\subsubsection{MAL-100}
In this paper, we paid significant effort to build and propose a new large-scale malware dataset named MAL-100. This dataset fills the gap of lacking large malware open-set benchmark dataset malware recognition domain (The MAL-100 will be released soon). It consists of more than 50 thousand malware instances covering a wide range of 100 malware families. 
To obtain the dataset, we first collected over 50 thousand PE malware files from VirusSign \cite{VirusSign}. Next, we applied VirusTotal \cite{VirusTotal} and AVClass \cite{sebastian2016avclass} to identify and label the malware families of the collected PE files. Then, the Ember \cite{anderson2018ember} is utilized to obtain the characteristics of each malware. 
Consequently, the malware characteristics used in our study consist of the following eight groups of raw characteristics including basic information from the PE header, information from the COFF header and optional header, imported functions, exported functions, section information, byte histogram, byte-entropy histogram, and printable-string information \cite{anderson2018ember}. Each malware instance is resized to a 25$\times$25 malware feature image. 
To have a deeper view on our proposed MAL-100, we calculate the mean-instances for each malware family and visualize the 100 families as grey-scale images. As demonstrated in Fig. 4, each image denotes one malware family in the $10 \times 10$ grid, and some more insights can be observed from the visualization results: 1) Each malware family contains sufficient features and thus can be applied to various models; 2) There exists more statistical characteristics and variances among different malware families which means that our MAL-100 has better representation properties and and can be used to train better malware recognition models; and 3) Our MAL-100 has great potential to be used in various deep learning techniques based on the image representation. In summary, our MAL-100 has a great potential and extension ability that contributing the future open-set malware research.


\subsection{Evaluation Criteria}
The performance of open-set malware recognition can be evaluated by two parallel tasks including the multi-families classification and the unknown detection.

\subsubsection{Multi-family Classification}
The multi-family classification aim at classifying malware instances belonging to known families as accurately as possible. The performance is evaluated by the classification accuracy $C_{Acc}$ defined as:
\begin{equation}
C_{Acc} = \frac{N_{correct}}{N_{instance}},
\end{equation}
where $N_{instance}$ is the total number of testing malware instances and $N_{correct}$ is the number of malware instances that are correctly classified. This definition corresponds to the conventional accuracy of most malware classification models.

\subsubsection{Unknown Detection}
The unknown detection aims at detecting malware instances belonging to novel unknown families with high accuracy rate.
To calculate the detection accuracy $D_{Acc}$, we need to consider the malware instances belonging to both known and unknown families during testing. We first calculate the true positive rate (TPR) for the instances of known families as:
\begin{equation}
	TPR_{(K)} = \frac{TP_{(K)}}{TP_{(K)}+FN_{(K)}},
\end{equation}
where $TP_{(K)}$ and $FN_{(K)}$ are the true positive and the false negative for known families, respectively, and then to calculate the true negative rate (TNR) for the instances of unknown families as:
\begin{equation}
	TNR_{(U)} = \frac{TN_{(U)}}{TN_{(U)}+FP_{(U)}},
\end{equation}
where $TN_{(U)}$ and $FP_{(U)}$ are the true negative and the false positive for unknown families, respectively. 
The detection accuracy $D_{Acc}$ is then calculated as:
\begin{equation}
D_{Acc} = \frac{TPR_{(K)} + TNR_{(U)}}{2}.
\end{equation}
 
\subsubsection{Competitors Selection}
In the experiments, the selection of the competitors is based on the following criteria: 1) All of these competitors are published in the most recent years; 2) They cover a wide range of models; 3) All of these competitors are under the same evaluation criteria; and 4) They clearly represent the state-of-the-art of malware recognition domain.


\subsection{Implementation}
\subsubsection{Network and Training Details}
Our model is implemented by Pytorch and trained on one GTX 1080Ti GPU. The classification network is built upon a 13-layer convolutional neural networks with ReLU activation and MaxPooling layers. The novelty synthesizing network is built on discriminator and one generator with 4-layer convolutional neural network, each. 
We specify the main structure of the classification network in TABLE I. 
As to the training details, the prior used for the generator is the normal (Gaussian) distribution with median 0 and variance 1. We use Adam optimizer for all networks including the classifier, discriminator, and generator. The learning rate is set to 0.0002, and the training runs for total 500 rounds.

\begin{table}[t]
\begin{center}
\caption{Main Structure of the Classification Network}
\label{table1}
\begin{tabular}{cccccc}        
\hline                   
Layer   &Channel        &Kernel       &Stride    &Padding     & RelU   \\
\hline
1-2     &32             &3$\times$3   & 1, 1     & 1, 1       & $\checkmark$  \\
3-4     &64             &3$\times$3   & 1, 1     & 1, 1       & $\checkmark$  \\
\hline 
\multicolumn{6}{c}{MaxPooling with kernel 2$\times$2, ktride 2, 2, dilation 1 } \\
\hline 
5     &64             &3$\times$3   & 1, 1     & 1, 1       & $\checkmark$  \\
6-7   &128             &3$\times$3   & 1, 1     & 1, 1       & $\checkmark$  \\
\hline 
\multicolumn{6}{c}{MaxPooling with kernel 2$\times$2, ktride 2, 2, dilation 1} \\
\hline 
8   &128             &3$\times$3   & 1, 1     & 1, 1       & $\checkmark$  \\
9-10   &256             &3$\times$3   & 1, 1     & 1, 1       & $\checkmark$  \\
\hline 
\multicolumn{6}{c}{MaxPooling with kernel 2$\times$2, ktride 2, 2, dilation 1} \\
\hline 
11   &256             &3$\times$3   & 1, 1     & 1, 1       & $\checkmark$  \\
12-13   &512             &3$\times$3   & 1, 1     & 1, 1       & $\checkmark$  \\
\hline 
\multicolumn{6}{c}{MaxPooling with kernel 2$\times$2, ktride 2, 2, dilation 1} \\
\hline
\multicolumn{6}{c}{Fully Connected layers} \\
\hline 
\end{tabular}
\end{center}
\end{table}

\begin{figure*}[t]
  \centering
  \subfigure[Accuracy during training]{
    \label{fig:subfig:a} 
    \includegraphics[width=2.7in]{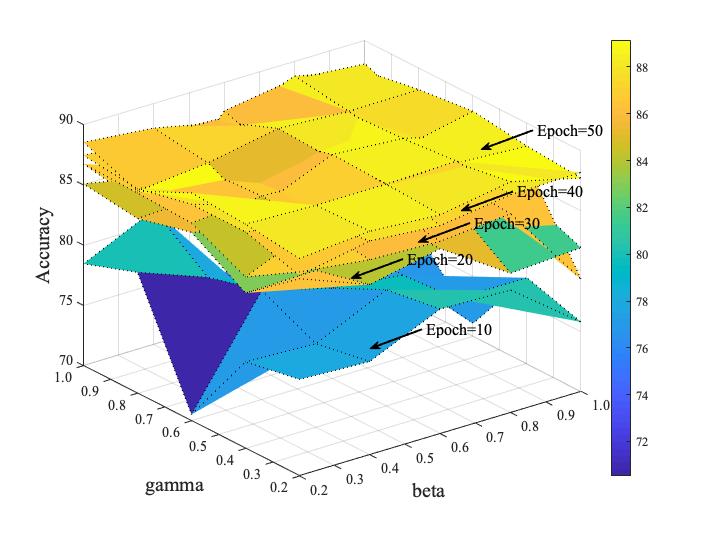}}\hspace{1cm}  
  \subfigure[Loss during training]{
    \label{fig:subfig:b} 
    \includegraphics[width=2.7in]{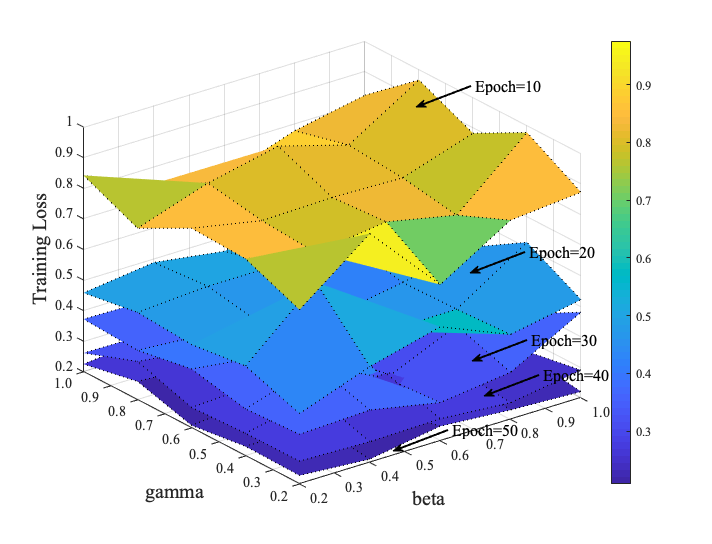}}
\caption{The sensibilities of hyper-parameters $\beta$ and $\gamma$: (a) effects on testing accuracy during training and (b) effects on loss during training (better viewed in color and zoom-in mode).}
  \label{beta_gamma}
\end{figure*}

\subsubsection{Dataset Preparation}
In our experiment, we conduct a thorough comparison to as many representative malware recognition models together with several image open-set recognition models as possible, to fully evaluate the effectiveness of our proposed method. Since most of the competitors are not open-sourced and the data splittings vary to a certain degree in different competitors, we take the following rules to prepare the experimental data. First, for the competitors whose source code is open-sourced or easily to be reproduced, we set the same data splitting to them and our proposed method, which is more strict and sacrificial (e.g., less training data v.s. more testing data) compared with other competitors. Second, for the competitors that are not easy to be reproduced with non-explicit training details, we fairly take their reported data splittings together with the evaluation results as the baselines in our experiment. Third, since the proposed MAL-100 dataset is new to all competitors, we thus can keep the consistency on the data splitting of MAL-100. In our experiment, the data splitting for each dataset are specified as follows.
\begin{itemize}
\item \textbf{BIG 2015}: The first 7 families containing 8,627 malware instances are selected as the known families, and the remaining 2 families containing 2,241 instances are selected as the novel unknown families for detection. For the known families, we randomly select about 80\%, i.e., 6,900, of the instances as the train-set, and the remaining 20\%, i.e., 1,727, are the test-set. Since the pre-processed instance vector consists of 622-dimensional features, hence we fill in three 0 at the end of each instance vector and resize it to a 25$\times$25 malware instance image to fit our model. For those competitors that do not require malware instance image input, we fed the 622-dimensional vector to them only. 
\item \textbf{Mailing}: The first 15 families containing 8,017 malware instances are selected as the known families, and the remaining 10 families containing 1,322 malware instances are selected as the novel unknown families for detection. Among the known families, about 80\%, i.e., 6400, of the instances are randomly selected as the train-set, and the remaining 20\%, i.e., 1617, are the test-set. Similarly, since the original bit grayscale malware images are with different size, so we also resize them into a uniform size of 32$\times$32. For those competitors that do not require malware instance image input, we resize them into 1-d vectors and fed them properly.
\item \textbf{MAL-100}: According to the ID order of total 100 families, the first 80 families containing 39,346 malware instances are selected as the known families, and the remaining 20 families containing 17,135 malware instances are selected as the novel unknown families for detection. 
For the known families, 80\%, i.e., 31,523 out of 39,346, of the instances are selected as the train-set, and the remaining 20\%, i.e., 7,823, are the test-set. Similarly, for those competitors that do not require malware instance image input, we resize them into 1-d vectors and fed them properly. 
\end{itemize}

It is worth noting that since our proposed synthesizing network is conditioned on known malware families to conservatively synthesize marginal malware instances to mimic the novel unknown malware families, we can thus easily have the understanding or observation that the more known malware families available usually result in better synthesized mimical unknown malware instances, which can eventually help to obtain better recognition performance especially for the detection task. Without loss of generality, we set 2,241 instances (out of a total 10,868) as unknown families for BIG 2015, 1,322 instances (out of a total 9,339) as unknown families for Mailing, and 17,135 instances (out of a total 56,481) unknown families for MAL-100.

\subsubsection{Hyper-parameters}
Two hyper-parameter $\beta$ and $\gamma$ are presented in Eq. (8) to balance the two rectification regularizers. Based on their definitions, $\beta$ and $\gamma$ control the unknown probabilities flattening and the known families exclusion in global and local conditions, respectively. These two rectification regularizers should be of equal importance, relatively and theoretically. To determine the importance of each term and set proper values to them, we briefly conduct a grid search to evaluate their sensibilities. Specifically, we set $\beta$ and $\gamma$ both range within [0.2-1.0] with an interval of 0.2, and run our proposed MOSR model on the MAL-100 dataset with only 50 epochs. We record the testing accuracy and training loss every 10 epochs to show the sensibilities. 

As demonstrated in Fig. \ref{beta_gamma}, we can observe that as the training move on, the testing accuracy and training loss gradually become indistinguishable with different $\beta$/$\gamma$ pairs, especially for the testing accuracy. The grid search proves that these two regularizers are of equal importance in rectifying the classifier. Hence, in our experiment, $\beta$ and $\gamma$ are both fixed to 1 without loss of generality. 

\begin{table*}[t]
\centering
\setlength{\tabcolsep}{4.2mm}{
\begin{threeparttable}  
\caption{Comparison Results on BIG 2015}  
\label{table2}    
\begin{tabular}{lcccccccc}  
\toprule  
\multirow{2}{*}{Method}
&\multirow{2}{*}{Group}
&\multicolumn{2}{c}{Known}
&\multicolumn{1}{c}{Unknown}
&\multicolumn{1}{c}{Classification}
&\multicolumn{1}{c}{Detection}\cr  

\cmidrule(lr){3-4} 
&    &Train \#  &Test \#             &Test \#            &$C_{Acc}$ (\%)      &$D_{Acc}$ (\%)  \cr  
\midrule  
Drew \textit{et al.} \cite{drew2016polymorphic}   &CML    &9,782       &1,086     & -         &97.42      & -      \cr
Narayanan \textit{et al.} \cite{narayanan2016performance}    &CML         &9,782       &1,086     & -    &96.60      & -      \cr
Burnaev \textit{et al.} \cite{burnaev2016one}     &CML         &6,900       &1,727     &2,241         &92.00      &49.75      \cr
Drew \textit{et al.} \cite{drew2017polymorphic}   &CML    &9,782       &1,086     & -         &98.59      & -      \cr
GISTSVM \cite{kalash2018malware}  &CML         &9,776       &1,092     & -         &88.74      & -      \cr
OpenMax \cite{bendale2016towards}    &DL         &6,900       &1,727     &2,241         &98.73      &78.40      \cr
tGAN \cite{kim2017malware}    &DL         &8,937       &997     & -         &96.39      & -      \cr
Kim \textit{et al.} \cite{kim2017image}    &DL         &9,782       &1,086     & -         &91.76      & -      \cr
Rahul \textit{et al.} \cite{rahul2017deep}    &DL         &9,782       &1,086     & -         &94.91      & -      \cr
SoftMax-DNNs     &DL         &6,900       &1,727     &2,241         &97.22      &67.30   \cr
tDCGAN \cite{kim2018zero}    &DL         &9,720       &1,080     & -        &95.74      & -      \cr
MCNN \cite{kalash2018malware}     &DL         &9,776       &1,092     & -         &98.99      & -      \cr
EVM \cite{rudd2018extreme}    &EVT         &6,900       &1,727     &2,241         &98.38      &73.18      \cr
\textbf{Ours}    &DL        &6,900       &1,727     &2,241         &\textbf{99.20}      &\textbf{92.77}      \cr

\bottomrule  
\end{tabular} 
\footnotesize{CML denotes the classic machine leaning-based method, DL denotes the deep leaning-based method, and ETV denotes the extreme value theory-based method; \# is the number of instances; '-' represents that there is no reported result.}
\end{threeparttable}
}
\end{table*} 

\subsection{Evaluation on BIG 2015}

\subsubsection{Competitors}
For BIG 2015 dataset, we compare our proposed method with 13 competitors including five classic machine leaning-based methods as Drew \textit{et al.} \cite{drew2016polymorphic}, Narayanan \textit{et al.} \cite{narayanan2016performance}, Drew \textit{et al.} \cite{drew2017polymorphic}, Burnaev \textit{et al.} \cite{burnaev2016one}, GISTSVM \cite{kalash2018malware}; seven deep learning-based methods as OpenMax \cite{bendale2016towards}, tGAN \cite{kim2017malware}, tDCGAN \cite{kim2018zero}, Kim \textit{et al.} \cite{kim2017image}, Rahul \textit{et al.} \cite{rahul2017deep}, SoftMax-DNNs, MCNN \cite{kalash2018malware}; and one extreme value theory-based method as EVM \cite{rudd2018extreme}.

\subsubsection{Comparison Results and Analysis}
The comparison results on BIG 2015 dataset is demonstrated in TABLE II. It can be observed that our model outperforms all competitors with the highest classification and detection (if applicable by competitors) accuracy as 99.20\% and 92.77\%, respectively. Moreover, two more observations can also be made. First, although our split of the dataset sacrifices more training data to be utilized as the testing data (both known and unknown), i.e., only 6,900 instances correspond about 63\% of the whole dataset for training, our model can still obtain the best results. Second, we can also note that our model improves the detection accuracy by a large margin as 14.37\%, which fully demonstrate the effectiveness.

\begin{table*}[t]
\centering
\setlength{\tabcolsep}{4.2mm}{
\begin{threeparttable}  
\caption{Comparison Results on Mailing}  
\label{table2}    
\begin{tabular}{lcccccccc}  
\toprule  
\multirow{2}{*}{Method}
&\multirow{2}{*}{Group}
&\multicolumn{2}{c}{Known}
&\multicolumn{1}{c}{Unknown}
&\multicolumn{1}{c}{Classification}
&\multicolumn{1}{c}{Detection}\cr  

\cmidrule(lr){3-4} 
&    &Train \#  &Test \#             &Test \#            &$C_{Acc}$ (\%)      &$D_{Acc}$ (\%)  \cr  
\midrule  
Nataraj \textit{et al.} \cite{nataraj2011malware}     &CML         &8,394       &945     & -         &97.18      & -  \cr
Burnaev \textit{et al.} \cite{burnaev2016one}     &CML         &6,400       &1,617     &1,322         &94.19      &52.72      \cr
Kalash \textit{et al.} \cite{kalash2018malware}     &CML         &8,394       &945     & -         &93.23      & -      \cr
Yajamanam \textit{et al.} \cite{yajamanam2018deep}    &CML      &8394       &945     & -         &97.00      & -    \cr
GIST+SVM \cite{cui2018detection}    &CML         &8,394       &945     & -         &92.20      & -      \cr
GLCM+SVM \cite{cui2018detection}    &CML         &8,394       &945     & -         &93.20      & -      \cr
SoftMax-DNNs     &DL         &6,400       &1,617     &1,322         &98.08      &73.45   \cr
tGAN \cite{kim2017malware}    &DL         &8,394       &945     & -        &96.82      & -      \cr
VGG-VeryDeep-19 \cite{yue2017imbalanced}    &DL         &5,603       &1,868     & -         &97.32      & -      \cr
Cui \textit{et al} \cite{cui2018detection}    &DL         &8,394       &945     & -         &97.60      & -      \cr
OpenMax \cite{bendale2016towards}    &DL         &6,400       &1,617     &1,322         &98.70      &84.34      \cr
IDA+DRBA \cite{cui2018detection}    &DL         &8,394       &945     & -         &94.50      & -      \cr
NSGA-II \cite{cui2019malicious}    &DL         &8,394       &945     & -         &97.60      & -      \cr
MCNN \cite{kalash2018malware}     &DL         &8,394       &945     & -         &98.52      & -      \cr
tDCGAN \cite{kim2018zero}    &DL         &8,394       &945     & -        &97.66      & -      \cr
IMCFN \cite{vasan2020imcfn}    &DL         &6,537       &2,802     & -         &98.82      & -      \cr
EVM \cite{rudd2018extreme}    &EVT         &6,400       &1,617     &1,322         &98.40      &81.70      \cr
\textbf{Ours}    &DL       &6,400       &1,617     &1,322         &\textbf{99.01}      &\textbf{88.90}      \cr
\bottomrule  
\end{tabular} 
\footnotesize{CML denotes the classic machine leaning-based method, DL denotes the deep leaning-based method, and ETV denotes the extreme value theory-based method; \# is the number of instances; '-' represents that there is no reported result.}
\end{threeparttable}
}
\end{table*}

\begin{table*}[t]
\centering
\setlength{\tabcolsep}{4.2mm}{
\begin{threeparttable}  
\caption{Comparison Results on MAL-100}  
\label{table2}    
\begin{tabular}{lcccccccc}  
\toprule  
\multirow{2}{*}{Method}
&\multirow{2}{*}{Group}
&\multicolumn{2}{c}{Known}
&\multicolumn{1}{c}{Unknown}
&\multicolumn{1}{c}{Classification}
&\multicolumn{1}{c}{Detection}\cr  

\cmidrule(lr){3-4} 
&    &Train \#  &Test \#             &Test \#            &$C_{Acc}$ (\%)      &$D_{Acc}$ (\%)  \cr  
\midrule
Burnaev \textit{et al.} \cite{burnaev2016one}     &CML         &31,523       &7,823     &17,135         &32.56      &52.58      \cr
SoftMax-DNNs     &DL         &31,523       &7,823     &17,135         &76.45      &63.02   \cr  
OpenMax \cite{bendale2016towards}    &DL         &31,523       &7,823     &17,135         &85.38      &70.57      \cr
tGAN \cite{kim2017malware}    &DL         &31,523       &7,823     &17,135         &77.23      &65.25      \cr
tDCGAN \cite{kim2018zero}    &DL         &31,523       &7,823     &17,135         &81.45      &67.04      \cr
EVM \cite{rudd2018extreme}    &EVT         &31,523       &7,823     &17,135         &84.10      &68.69      \cr
\textbf{Ours}    &DL        &31,523       &7,823     &17,135         &\textbf{91.17}      &\textbf{86.23}      \cr
\bottomrule  
\end{tabular} 
\footnotesize{CML denotes the classic machine leaning-based method, DL denotes the deep leaning-based method, and ETV denotes the extreme value theory-based method; \# is the number of instances.}
\end{threeparttable}
}
\end{table*}


\subsection{Evaluation on Mailing}

\subsubsection{Competitors}
For Mailing dataset, we compare our proposed method with 17 competitors including six classic machine leaning-based methods as Nataraj \textit{et al.} \cite{nataraj2011malware}, Kalash \textit{et al.} \cite{kalash2018malware}, Yajamanam \textit{et al.} \cite{yajamanam2018deep}, Burnaev \textit{et al.} \cite{burnaev2016one}, GIST+SVM \cite{cui2018detection}, GLCM+SVM \cite{cui2018detection}; ten deep learning-based methods as Vgg-verydeep-19 \cite{yue2017imbalanced}, Cui \textit{et al} \cite{cui2018detection}, IDA+DRBA \cite{cui2018detection}, NSGA-II \cite{cui2019malicious}, MCNN \cite{kalash2018malware}, IMCFN \cite{vasan2020imcfn}, OpenMax \cite{bendale2016towards}, tGAN \cite{kim2017malware}, tDCGAN \cite{kim2018zero}; and one extreme value theory-based method as EVM \cite{rudd2018extreme}.

\subsubsection{Comparison Results and Analysis}
The comparison results on Mailing dataset is demonstrated in TABLE III. Again, we can observe that our model also obtains the best results in both classification and detection (if applicable by competitors) accuracy as 99.01\% and 88.90\%, respectively. Moreover, the detection accuracy is also improved with a margin as 4.56\%.




\subsection{Evaluation on MAL-100}
\subsubsection{Competitors}
For our proposed MAL-100 dataset, we compare our proposed model with 6 competitors including one classic machine leaning-based method as Burnaev \textit{et al.} \cite{burnaev2016one}; four deep learning-based methods as OpenMax \cite{bendale2016towards}, SoftMax-DNNs, tGAN \cite{kim2017malware}, tDCGAN \cite{kim2018zero}; and one extreme value theory-based method as EVM \cite{rudd2018extreme}. Specifically, we adopt the same dataset split for all competitors and our model for a better demonstration of our proposed MAL-100 dataset.

%

\subsubsection{Comparison Results and Analysis}
The comparison results on MAL-100 dataset is demonstrated in TABLE IV. It can be observed that our model outperforms these competitors by great advantages with the classification and detection accuracy as 91.17\% and 86.23\%, respectively. The improved margin are 5.79\% in classification task, and 15.66\% for detection task. 
Among them, the classic machine leaning-based method, i.e., Burnaev \textit{et al.} \cite{burnaev2016one}, has a fairly good performance in detection accuracy as 52.58\%, which contrasts with the poor performance of its classification ability. The performance of deep learning-based methods, i.e., SoftMax-DNNs, OpenMax \cite{bendale2016towards}, tGAN \cite{kim2017malware} and tDCGAN \cite{kim2018zero} are relatively stable for the detection accuracy, where they achieve 63.02\%, 70.57\%, 65.25\% and 67.04\%, respectively. 
The extreme value theory-based method, i.e., EVM, obtains the sub-optimal performance with a detection accuracy of 68.69\%. Again, our model obtains the best result in detection with an accuracy of 86.23\%, which is far exceeds other competitors and expands the advantages of our model. 

\begin{figure*}[t]
\centering
\includegraphics[width=0.7\textwidth]{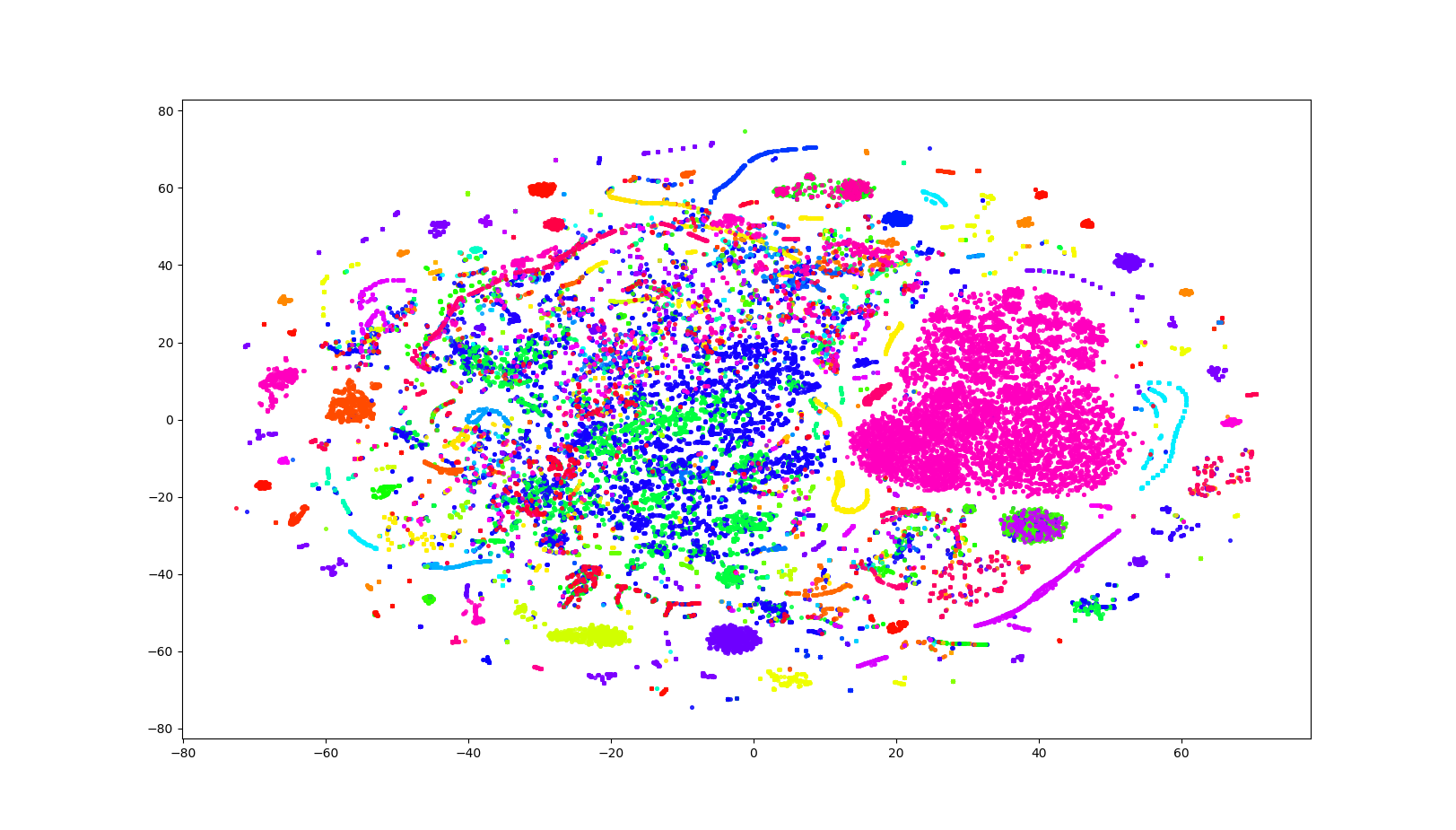}
\caption{Full visualization of our MAL-100 dataset by tsne \cite{maaten2008visualizing}. It can be observed that there exists many overlaps among different malware families, which makes a comparatively smaller variances compared with images domain (better viewed in color).}
\label{tsne_mal100}
\end{figure*}

It is worth noting that, as demonstrated in Fig. \ref{tsne_mal100}, since the proposed MAL-100 is a large-scale malware dataset with complex data distribution and more overlaps among different malware families. Existing methods can hardly distinguish various malware families and result in mediocre results, especially for the detection accuracy. In contrast, our model can mitigate this problem by conservatively synthesizing several marginal malware instances to mimic the novel unknown families and support the training of the classifier. Thus, our model can still obtain a superior and balanced performance in both classification and detection tasks.

%
%
%

\begin{table}[t]
\centering
\begin{threeparttable}  
\caption{Ablation Study}    
\begin{tabular}{cccccc}  
\toprule  
\multicolumn{4}{c}{Module}
&\multicolumn{1}{c}{Classification}
&\multicolumn{1}{c}{Detection}\cr  

\cmidrule(lr){1-4} 
$(i)$    &$(ii)$  &$(iii)$  &$(iv)$  &$C_{Acc}$ (\%)  &$D_{Acc}$ (\%)      \cr  
\midrule  
$\checkmark$   & & &    &87.09  &71.42       \cr
$\checkmark$   &$\checkmark$ & &$\checkmark$   &88.57  &82.69       \cr
$\checkmark$   & &$\checkmark$ &$\checkmark$   &89.28  &83.66       \cr
$\checkmark$   &$\checkmark$ &$\checkmark$ &$\checkmark$ &91.17  &86.23       \cr
\bottomrule  
\end{tabular} 
\end{threeparttable}
\end{table}

\subsection{Ablation Analysis}
To further confirm the utility and effectiveness of our proposed method, we conduct an ablation analysis on the unified training objective in four scenarios specified from Eq. (8): 1) Naive classification network with only supervision $(i)$ (i.e., cross-entropy); 2) Classification network trained with supervision $(i)$, $(ii)$, and $(iv)$ (i.e., with synthesizing process and known families exclusion regularizer); 3) Classifier trained with supervision $(i)$, $(iii)$, and $(iv)$ (i.e., with synthesizing process and unknown probabilities flattening regularizer); and 4) Classifier trained with full supervision $(i)$, $(ii)$, $(ii)$, and $(iv)$. Specifically, each scenario is performed in a full training process on MAL-100 dataset with 500 rounds. 

The comparison results are demonstrated in TABLE V. It can be seen from the results that the native classifier in our proposed method performs a fairly acceptable classification accuracy as 87.09\%, while a relatively low detection accuracy as 71.42\%. By adding the synthesizing process with rectification regularizers, i.e., known families exclusion regularizer $(iii)$ or unknown probabilities flattening regularizer $(ii)$, we can improve the recognition performance especially for the detection accuracy, where the margins are 11.27\% and 12.24\%, respectively. This difference indicates that the unknown probabilities flattening regularizer performs slightly better than that of the known families exclusion regularizer. Surprisingly, combining the synthesizing process with both rectification regularizers, we can obtain a significant improvement on both the classification and detection accuracy as 91.17\% and 86.23\%, respectively, which fully demonstrates the utility and effectiveness of our method.

\begin{figure*}[t]
  \centering
  \subfigure[Classification confusion matrix on MAL-100]{
    \label{fig:subfig:a} 
    \includegraphics[width=2.8in]{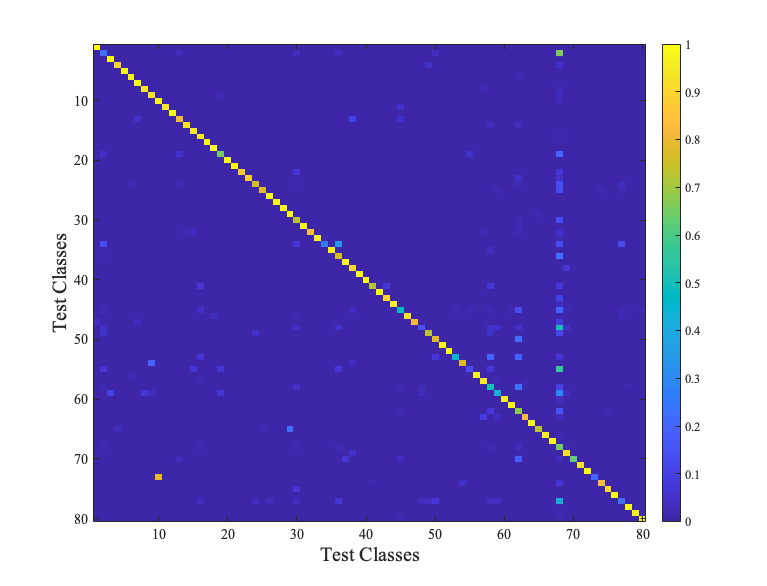}}\hspace{1cm}  
  \subfigure[Landscape of the confusion matrix]{
    \label{fig:subfig:b} 
    \includegraphics[width=2.8in]{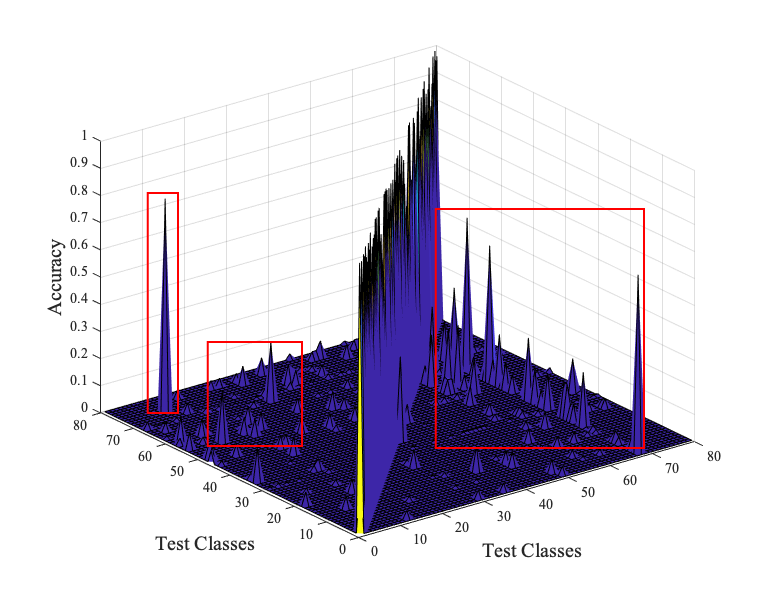}}
\caption{Per-family classification performance on MAL-100: (a) the classification confusion matrix and (b) the landscape of the the confusion matrix. Some hard malware families can be observed as family ``1'', ``33'', ``44'', ``47'', ``54'', ``58'', ``72'', and so on (better viewed in color and zoom-in mode).}
  \label{CM}
\end{figure*}

\subsection{Fine-grained Accuracy}

To further evaluate the predictive power of our method and explore the hard malware families of MAL-100 (i.e., malware families that are easily misclassified), we record and count the prediction results for each testing malware families and analyze the fine-grained per-family classification performance. This evaluation is conducted on the proposed MAL-100 dataset with a full training process. 

As shown in Fig. \ref{CM}, the per-family classification results are presented by the confusion matrix and its corresponding landscape. In the confusion matrix, the column position indicates the ground truth, and the row position denotes the predicted results. The diagonal position thus indicates the classification accuracy for each malware family. From the results we can observe that our method can obtain quite good performance on MAL-100, which demonstrates its effectiveness. Despite the superior overall performance, we can also observe some hard malware families in MAL-100. For example, family ``1'' (ID starts from 0) is misclassified to family ``67'' with the probability as 65.63\% (i.e., denoted as family ``1'' $\rightarrow$ family ``67'': 78.46\%). Similarly, we list some more hard malware families including family ``33'' $\rightarrow$ family ``35'': 31.25\%; family ``44'' $\rightarrow$ family ``67'': 19.05\%; family ``47'' $\rightarrow$ family ``67'': 50.00\%; family ``54'' $\rightarrow$ family ``67'': 56.25\%; family ``58'' $\rightarrow$ family ``67'': 28.13\%; and family ``72'' $\rightarrow$ family ``9'': 78.46\%. Among them, family ``72'' becomes the hardest malware family in MAL-100 and the results suggest that we can pay more attention to the testing malware instances that are predicted to family ``9'', to double-check whether is from family ``72''. Moreover, we can surprisingly observe that most hard malware families of MAL-100 are quite commonly misclassified to family ``67'' (i.e., 5 out of 7 hardest malware families). Such results can suggest that family ``67'' may contain more common characteristic features that a majority of malware families have, and a ``focus list'' for those malware instances that are predicted as family ``67'' can be established to double-check the correctness. 

\begin{figure*}[t]
  \centering
  \subfigure[Synthesized Families v.s. Known Families]{
    \label{fig:subfigSV:a} 
    \includegraphics[width=2.8in]{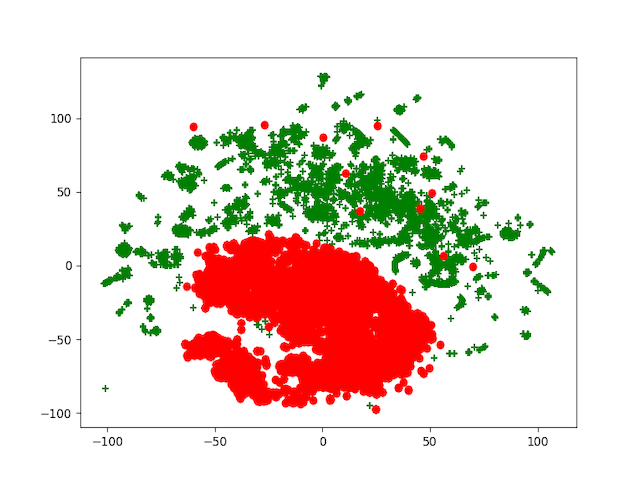}}\hspace{1cm}  
  \subfigure[Synthesized Families v.s. Unknown Families]{
    \label{fig:subfigSV:b} 
    \includegraphics[width=2.8in]{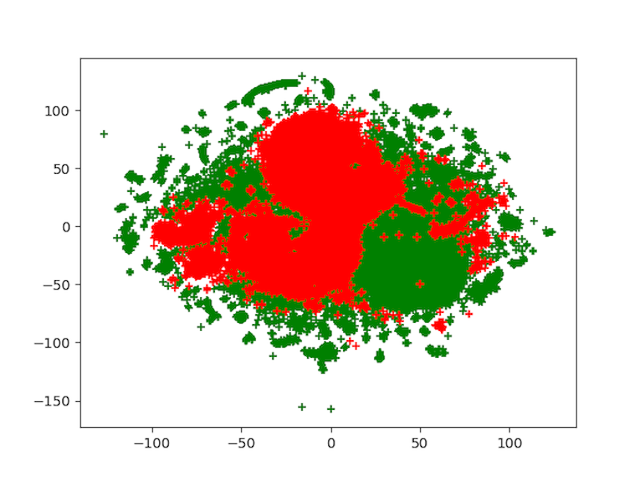}}
\caption{The visualization results of the synthesizing network on MAL-100: (a) synthesized families (denoted in red) v.s. known families (denoted in green) and (b) synthesized families (denoted in red) v.s. unknown families (denoted in green) (better viewed in color).}
  \label{SV}
\end{figure*}

\subsection{Further Analysis and Discussion}

\subsubsection{Synthesizing Visualization}
We visualize the synthesizing results of the trained synthesizing network in Fig. \ref{SV}. Specifically, our MOSR system is performed on a random subset of MAL-100 dataset to train the synthesizing network. In \ref{fig:subfigSV:a}, we present the tsne \cite{maaten2008visualizing} visualization results of $\sim7000$ instances from synthesized malware families (denoted as red) and the same number of instances from known malware families (denoted as green). We can observe that the synthesized malware families are clustered in the low-density region of known malware families, which indicates that the synthesized instances are close to known malware families while not belong to any of them, as expected. In \ref{fig:subfigSV:b}, we present the tsne visualization results of instances from synthesized malware families (denoted as red) and instances from unknown malware families (denoted as green). It can be observed from the results that a certain range of overlap exists between the synthesized and unknown malware families, which suggests an expected synthesizing ability of the trained synthesizing network.

\subsubsection{System Complexity}
Our method contains two main components including 1) the classification network that aims at producing accurate probability distributions for known malware families, and flat and low probability distributions of unknown ones, and 2) the novelty synthesizing network that aims to synthesize marginal malware instances of known families. Network 1) and 2) are jointly trained during the training phase, while network 1) is solely performed during the inference phase. Since the synthesizing network is implemented by the GANs that require numerous resources, the computation overhead of our proposed MOSR model is exponentially greater than that of most classic machine learning-based methods. However, since the total training rounds of our method are only 500, in contract to 2,000 rounds of tDCGAN \cite{kim2018zero}, 10,000 rounds of Lu \textit{et al.} \cite{lu2019generative}, etc., which also use the GANs, the training process of our method is more efficient than most GANs-based malware recognition models. 
On the other hand, the inference process of our method involves only one rectified 13-layer convolutional neural network, which is light yet more efficient than most existing malware recognition models range from classic machine learning-based methods to broad coverage of deep learning-based methods.

\section{Conclusion and Future Work}
\label{conclusion}
In this paper, we proposed a novel applicable malware recognition model in the open-set scenario. 
This model involves a synthesizer to conservatively synthesize marginal malware instances to mimic novel unknown families that can rectify the recognition performance of both classification and detection. 
Moreover, we also proposed a new large-scale malware dataset, named MAL-100, to fill the gap of lacking large open-set malware benchmark dataset and constantly contributes to the future research of malware recognition. 

In the future, we have two research routes to further improve the malware open-set recognition. The first route investigates the more efficient generative frameworks to synthesize malware instances for both the novelty synthesizing and data augmentation, which can then facilitate both the training and inference process. The second route focuses on going from the detection of unknown malware families to the classification of them. Such as target can be theoretically achieved by exploiting a shared semantic feature space between known and unknown malware families, of which the concept can be borrowed from the perspective of zero/few-shot learning. The challenge mainly lies in how each malware family can be semantically related to each other.


\ifCLASSOPTIONcaptionsoff
  \newpage
\fi

\bibliographystyle{IEEEtran}
\bibliography{my}
\end{document}